\documentclass[traditabstract]{aa}
\usepackage{txfonts}
\usepackage{natbib}
\usepackage{graphicx}
\usepackage{color}
\bibpunct{(}{)}{;}{a}{}{,} 
\usepackage{color}

\begin{document}

\title{Detectability of ultrahigh energy cosmic ray signatures\\ in gamma rays}

\author{Kumiko Kotera\inst{1,2}
  \and Denis Allard\inst{3}
     \and Martin Lemoine\inst{1}}

\offprints{\email{ kotera@uchicago.edu }}

\institute{Institut d'Astrophysique de Paris
UMR7095 - CNRS, Universit\'e Pierre \& Marie Curie,
98 bis boulevard Arago
F-75014 Paris, France.\and 
Department of Astronomy \& Astrophysics, Enrico Fermi
  Institute, and Kavli Institute for Cosmological Physics, The
  University of Chicago, Chicago, Illinois 60637, USA. \and
  Laboratoire Astroparticules et Cosmologie (APC), Universit\'e Paris
  7/CNRS, 10 rue A. Domon et L. Duquet, 75205 Paris Cedex 13, France.}

\authorrunning{K. Kotera et al.}
\titlerunning{Detectability of ultrahigh energy cosmic ray signatures in gamma rays}

\date{\today}

\abstract{The injection of ultra-high energy cosmic rays in the
  intergalactic medium leads to the production of a GeV-TeV gamma-ray
  halo centered on the source location, through the production of a
  high electromagnetic component in the interactions of the primary
  particles with the radiation backgrounds. This paper examines the
  prospects for the detectability of such gamma ray halos. We explore
  a broad range of astrophysical parameters, including the
  inhomogeneous distribution of magnetic fields in the large scale
  structure as well as various possible chemical compositions and
  injection spectra; and we consider the case of a source located
  outside clusters of galaxies. With respect to the gamma-ray flux associated 
  to synchrotron radiation of ultra-high energy secondary pairs, we 
  demonstrate that it does not depend strongly on these parameters and conclude that
  its magnitude ultimately depends on the energy injected in the
  primary cosmic rays. Bounding the cosmic ray luminosity with the
  contribution to the measured cosmic ray spectrum, we then find that the
  gamma-ray halo produced by equal luminosity sources is well below
  current or planned instrument sensitivities. Only rare and powerful
  steady sources, located at distances larger than several hundreds of
  Mpc and contributing to a fraction $\gtrsim 10\,$\% of the flux at
  $10^{19}\,$eV might be detectable. We also discuss the gamma-ray halos that are produced by inverse Compton/pair production cascades seeded by ultra-high energy cosmic rays. This latter signal strongly depends on the configuration of the extragalactic magnetic fields; it is dominated by the synchrotron signal on a degree scale if the filling factor of magnetic fields with $B\gtrsim10^{-14}$~G is smaller than a few percents. Finally, we discuss briefly the case of nearby potential sources such as Centaurus~A.}

\maketitle

\section{Introduction}

The quest for the sources of ultrahigh energy cosmic rays has long
been associated with the search of their secondary radiative
signatures. While propagating, the former indeed produce very high
energy photons through the interactions with the ambient backgrounds;
and gamma rays should be valuable tools for the identification of the
birthplace of the primary cosmic rays as they travel in a straight
manner, contrarily to charged particles.

The detection of such photon fluxes is far from straightforward,
however. On purely astrophysical grounds, the propagation of gamma
rays with energy exceeding several TeV is obstructed by their
relatively short pathlength of interaction with cosmic microwave
background (CMB) and infrared photons. These interactions lead to the
production of high energy electron and positron pairs which in turn
up-scatter CMB or radio photons by inverse Compton processes,
initiating electromagnetic cascades. One does not expect to observe
gamma rays of energy above $\sim 100$~TeV from sources located beyond
a horizon of a few megaparsecs
\citep{Gould67,WTW72,Protheroe86,PS93}. Below this energy, the
detection of the gamma ray signal originating from ultrahigh energy
cosmic rays depends on the luminosity and angular extension of the
observed object and obviously on the sensitivity and angular
resolution of the available instruments.

Intergalactic magnetic fields inevitably come into play in this
picture: i) primary cosmic rays can be deflected by the magnetic field
surrounding the source prior to the production of secondary
photons/electrons/positrons, ii) secondary electrons and positrons
produced by cosmic ray interactions or by photon-photon interactions
can also be deflected by the intergalactic magnetic field during the
cascade. In both cases the ultimate gamma ray emission can experience
a sizable spread around the ultrahigh energy cosmic ray
source. 

\cite{Aharonian02} and \cite{GA05} have observed that the magnetic
field environment of the source may also lead to the generation of a
multi-GeV gamma ray halo around ultra-high energy cosmic ray sources.
The idea is the following: for high enough energies and in the
presence of a magnetic field large enough in the surroundings of the
source, the electron and positron pairs produced through the
interaction of ultra-high energy particles with the ambient
backgrounds may lose a significant fraction of their energy through
synchrotron emission. This synchrotron component falls around
$\sim10$~GeV, below the pair creation threshold, hence it will not be
affected by further electromagnetic cascading. Assuming that the
source is located at 100~Mpc distance, at the center of a
homogeneously magnetized sphere of radius 20~Mpc with $B=1~$nG,
\cite{GA05} estimate that this signal would have an angular size of a
fraction of degree (see also \citealp{Aharonian10}).  It could therefore appear point-like for
instruments such as the Fermi Space Telescope but as an extended
source for future imaging Cerenkov telescopes. 

This process is particularly interesting, because its signature could
easily be differentiated from a point-like gamma ray signal emitted by
leptonic or hadronic channels inside the source, associated with the
acceleration of particles to energies possibly well below
$10^{19}\,$eV. The dominant background to this synchrotron halo
  is actually the halo produced by multi-TeV electrons that undergo
  inverse Compton on the cosmic microwave background and that were
  seeded by multi-TeV photons at $\gtrsim\,$Mpc distance from the
  source~\citep{ACV94}. However, as discussed in the
  present paper, discrimination between these two signals should be
  possible thanks to the different dependencies of the angular images
  on physical parameters and on energy. Then, one could argue that the
  detection of a synchrotron halo around a powerful source would
  constitute an unambiguous signature of acceleration to ultra-high
  energies in this source: first of all, the inverse Compton cooling
  length of electrons with energy $1\,{\rm TeV}\,\lesssim\,E_e\lesssim
  10^{17}\,$eV is smaller than $300\,$kpc, hence the gamma-ray signal
  of very high energy electrons produced in the source would appear
  point-like at distances $\gtrsim100\,$Mpc; furthermore, secondary
  electrons generally carry a few percents of the parent proton
  energy, so that $\gtrsim 10^{17}\,$eV secondary electrons correspond
  to $\gtrsim 10^{19}\,$eV protons; finally, if the radiating
  electrons are seeded away from the source by high energy nuclei,
  these nuclei must carry an energy $\gtrsim 10^{19}\,$eV as lower
  energy nuclei are essentially immune to radiative losses.

Given our state of knowledge on the sources of ultrahigh energy cosmic
rays, the detection of such a halo would have a lasting impact on this
field of research. The detectability of such signals thus deserves
close scrutiny.

The production of gamma ray signatures of ultrahigh energy cosmic rays
have been studied numerically by \cite{Ferrigno04} and
\cite{ASM06}. However, the former authors have focused their study on
the Compton cascades following the production of ultrahigh energy
photons and pairs close to the source and they have neglected the
deflection imparted by the surrounding magnetic fields which dilutes
the gamma-ray signal \citep{GA05}, while \cite{ASM06} have
addressed both the synchrotron emission of secondary pairs and the
Compton cascading down to TeV energies, albeit for the particular case
of a source located in a magnetized cluster of galaxies. The high
magnetic field that prevails in such environments increases the
residence time of primary and secondary charged particles and thus
also increases the gamma ray flux.

The present paper aims at examining the prospects for the
detectability of gamma ray halos around ultrahigh energy cosmic ray
sources, relaxing most of the assumptions made in the above previous
studies. In particular, we discuss the more general case of a source
located in the field, outside clusters of galaxies. We focus our
discussion on the synchrotron signal emitted by secondary pairs, which
offers a possibility of unambiguous detection; nevertheless, the
deflection and the dilution of the Compton cascading gamma ray signal
at TeV energies is also discussed. Going further than
\cite{Aharonian02} and \cite{GA05}, we take into account the
inhomogeneous distribution of the magnetic fields in the source
environment. We also relax the assumption of a pure proton composition
of ultra high energy cosmic rays, underlying to the above studies. The
chemical composition of ultrahigh energy cosmic rays indeed remains an
open question. While experiments such as the Fly's Eye and HiRes have
suggested a transition from heavy to light above $\sim 10^{18.5}$~eV
\citep{Fly, Hires, Hires10}, the most recent measurements made with
the Pierre Auger Observatory rather point towards a heavy composition
above $10^{19}\,$eV \citep{U07,Abraham09_comp,Auger10}.  As the energy
losses and magnetic deflection of high energy nuclei differ from those
of a proton of a same energy, one should naturally
expect different gamma ray signatures. 

The lay-out of the present paper is as follows. In
Section~\ref{section:B}, we first test the dependence of the gamma ray
flux produced by ultrahigh energy cosmic rays on the type, intensity
and structure of magnetized environments.  We also discuss the effects
of various chemical compositions and injection spectra. We conclude on
the robustness of the gamma ray signature according to these
parameters and find that the normalization and thus the detectability
of this flux ultimately depends on the energy injected in the primary
cosmic rays. In Section~\ref{section:detectability}, we discuss the
detectability of ultrahigh energy cosmic ray signatures in gamma
rays. Applying the results of our calculations, we show that the
average type of sources contributing to the ultrahigh energy cosmic
ray spectrum produces a gamma ray flux more than two orders of
magnitudes lower than the sensitivity of the current and upcoming
instruments. We then explore the case of rare powerful sources with
cosmic ray luminosity over energy $10^{19}$~eV of
$L_{19}>10^{44-46}$~erg$\,$s$^{-1}$. We assume throughout this paper that sources emit isotropically and discuss how the conclusions are modified for beamed emission in Section~\ref{section:conclusion}. The gamma ray signatures of those
sources could be detectable provided that they are located far enough
not to overshoot the observed cosmic ray spectrum. Finally, we also
briefly discuss the detection of nearby sources, considering the
radiogalaxy Centaurus~A as a protoypical example. We draw our
conclusions in Section~\ref{section:conclusion}.

\section{Simulations} \label{section:B}

As mentioned above, we focus our discussion on the synchrotron signal
that can be produced close to the source by very high energy electrons
and positrons, that result themselves from the interactions of primary
cosmic rays. The signal associated to inverse Compton cascades of
these electrons/positrons on radiation backgrounds will be discussed
in Section~\ref{sec:compton-cascade}.

The secondary electrons and positrons are created through one of the
three following channels: i) by the decay of a charged pion
produced during a photo-hadronic interaction ($A\,\gamma\rightarrow
\pi^+ +...$, then $\pi^+\rightarrow \mu^+ \, \nu_\mu$, and
$\mu^+\rightarrow e^+ \, \nu_e \, \bar{\nu_\mu}$, with $A$ a cosmic
ray nucleus), ii) by photo pair production during an interaction with
a background photon ($A\,\gamma\rightarrow e^+\,e^- +...$), or iii) by
the disintegration of a neutral pion into ultrahigh energy photons
which then interact with CMB and radio backgrounds to produce electron
and positron pairs ($A\,\gamma\rightarrow \pi^0 +...$, then
$\pi^0\rightarrow 2\gamma$, and $\gamma\,\gamma_{\rm bg}\rightarrow
e^+ \, e^-$, with $\gamma_{\rm bg}$ a cosmic background photon). In
all cases, the resulting electrons and positrons typically carry up to
a few percents of the initial cosmic ray energy.

While propagating in the intergalactic medium, these ultrahigh energy
electrons and positrons up-scatter CMB or radio photons through
inverse Compton processes and/or they lose energy through synchrotron
radiation. Following ~\cite{GA05}, the effective inverse Compton
cooling length\footnote{ Let us recall that at very high energy, pair
  production ($\gamma \gamma_{\rm bg}\rightarrow e^+e^-$) transfers
  energy to one of the pairs. The inverse Compton scattering of
  photons ($e\,\gamma_{\rm{bg}}\rightarrow e\,\gamma$) occurring in
  the Klein-Nishina regime, nearly all the electron or positron energy
  is again transfered to the up-scattered photon. Thus the initial
  electron energy in a cascade is degraded very slowly, until the
  electron or positron energy is either radiated in synchrotron, or
  the photon energy falls beneath the pair production threshold. For
  this reason one may consider the cascade ($e\rightarrow \gamma
  \rightarrow e ...$) as the disintegration of one single particle,
  that loses energy over an effective loss length $x_{e \gamma}$
  \citep{Stecker73,Gould78}.  } on the CMB and radio backgrounds can
be written as $x_{e\gamma}\,\approx\,5\,{\rm Mpc}\,(E_e/10^{18}\,{\rm
  eV})^{\alpha_{\rm IC}}$, with $\alpha_{\rm IC}=1$ if the electron
energy $E_e\lesssim 10^{18}\,$eV and $\alpha_{\rm IC}\simeq 0.25$ if
$10^{18}\,{\rm eV}\,\lesssim E_e\lesssim 10^{20}\,$eV. Above
$10^{18}$~eV, the scaling of the cooling length actually depends on
the assumptions made for the radio background, which is unfortunately
not very well known. The possible differences that such uncertainties
could introduce should however not affect the results discussed in
this paper. This length scale has to be compared with the synchrotron
cooling length
\begin{equation}
  x_{e B} \sim 3.8~{\rm kpc}\,\left(\frac{B}{10~{\rm nG}}\right)^{-2} 
  \left(\frac{E_e}{10^{19}~{\rm eV}}\right)^{-1}\ ,
\end{equation}
where $B$ is the magnetic field intensity (assumed homogeneous over
this distance) and $E_e$ the electron or positron energy. 

As discussed in \cite{GA05}, the opposite scalings of $x_{e\gamma}$
and $x_{eB}$ with electron energy imply the existence of a cross-over
energy $E_\times$, which depends on $B$ and is such that beyond
$E_\times$, electrons mainly cool via synchrotron instead of
undergoing an inverse Compton cascade. For a magnetic field of
intensity $B=10^{-9}B_{\rm nG}\,$G, $E_\times \sim 10^{18}\,{\rm
  eV}B_{\rm nG}^{-1}$ for $B_{\rm nG}\gtrsim 1$, $E_\times \sim
10^{18}\,{\rm eV}B_{\rm nG}^{-1.6}$ for $1\gtrsim\,B_{\rm nG}\gtrsim
0.1$ and $E_\times \sim 10^{20}\,{\rm eV}B_{\rm nG}^{-1}$ for
$0.1\gtrsim\,B_{\rm nG}$, see Fig.~1 of \cite{GA05} and see
\cite{Lee98} for further details on the cascade physics.

Finally, the emitted synchrotron photon spectrum peaks at the energy:
\begin{equation}\label{eq:Esyn}
  E_{\gamma,{\rm syn}} \sim 68 \,{\rm GeV}\,\left(\frac{B}{10~{\rm
        nG}}\right)
  \left(\frac{E_e}{10^{19}~{\rm eV}}\right)^2\, .
\end{equation}

\begin{figure*}
\centering
   \includegraphics[width=0.31\textwidth]{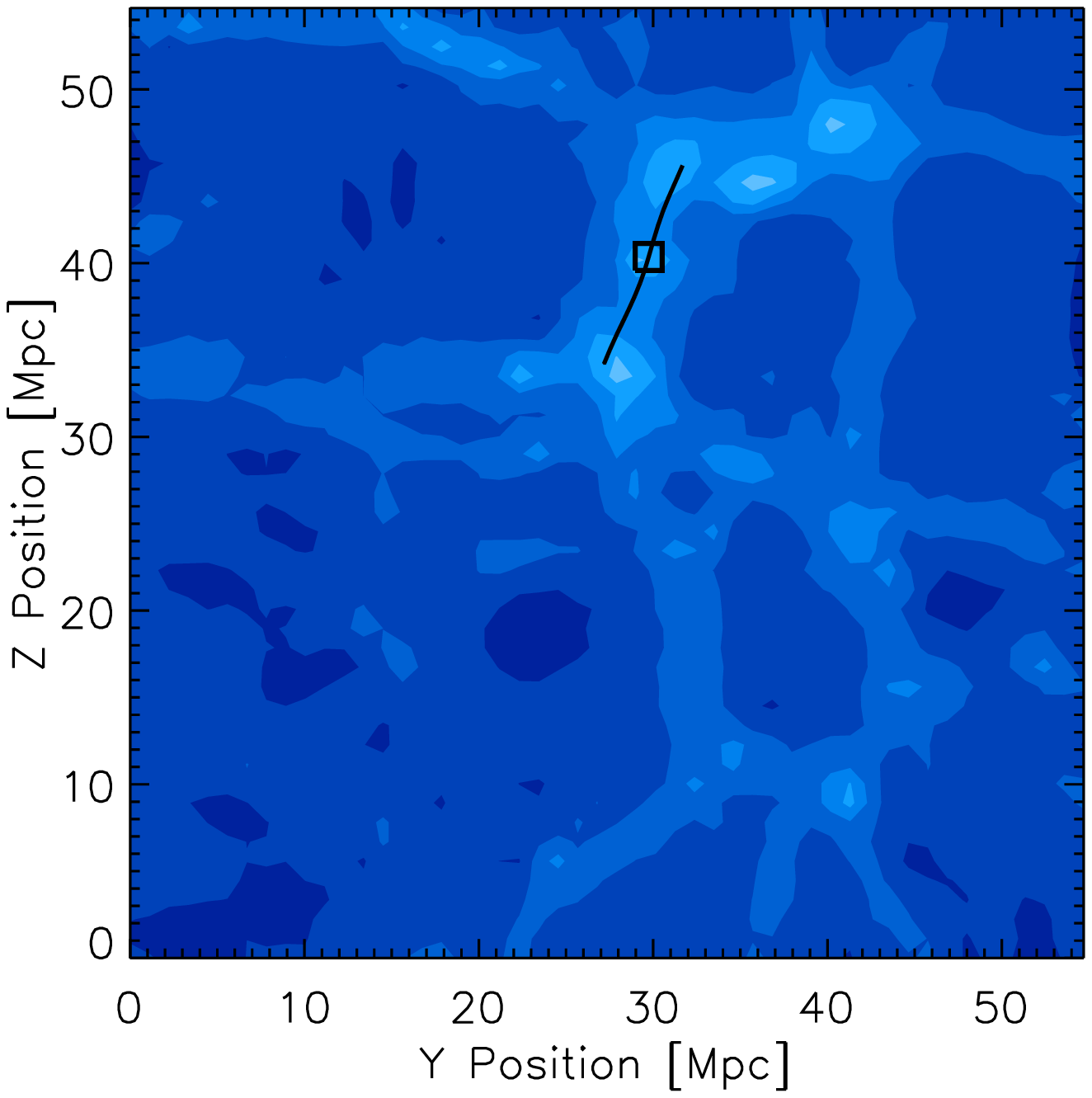}
   \includegraphics[width=0.31\textwidth]{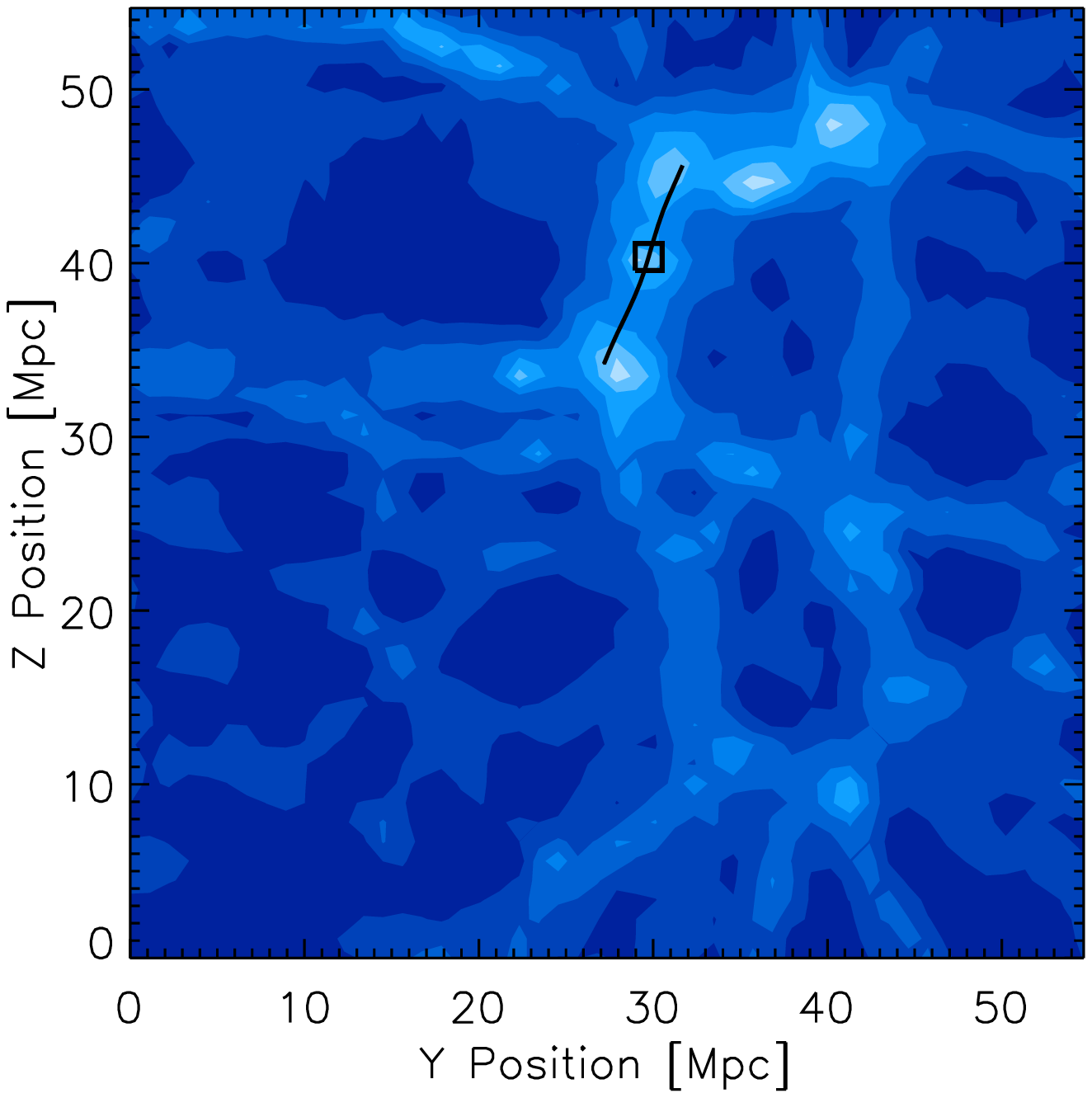}
   \includegraphics[width=0.362\textwidth]{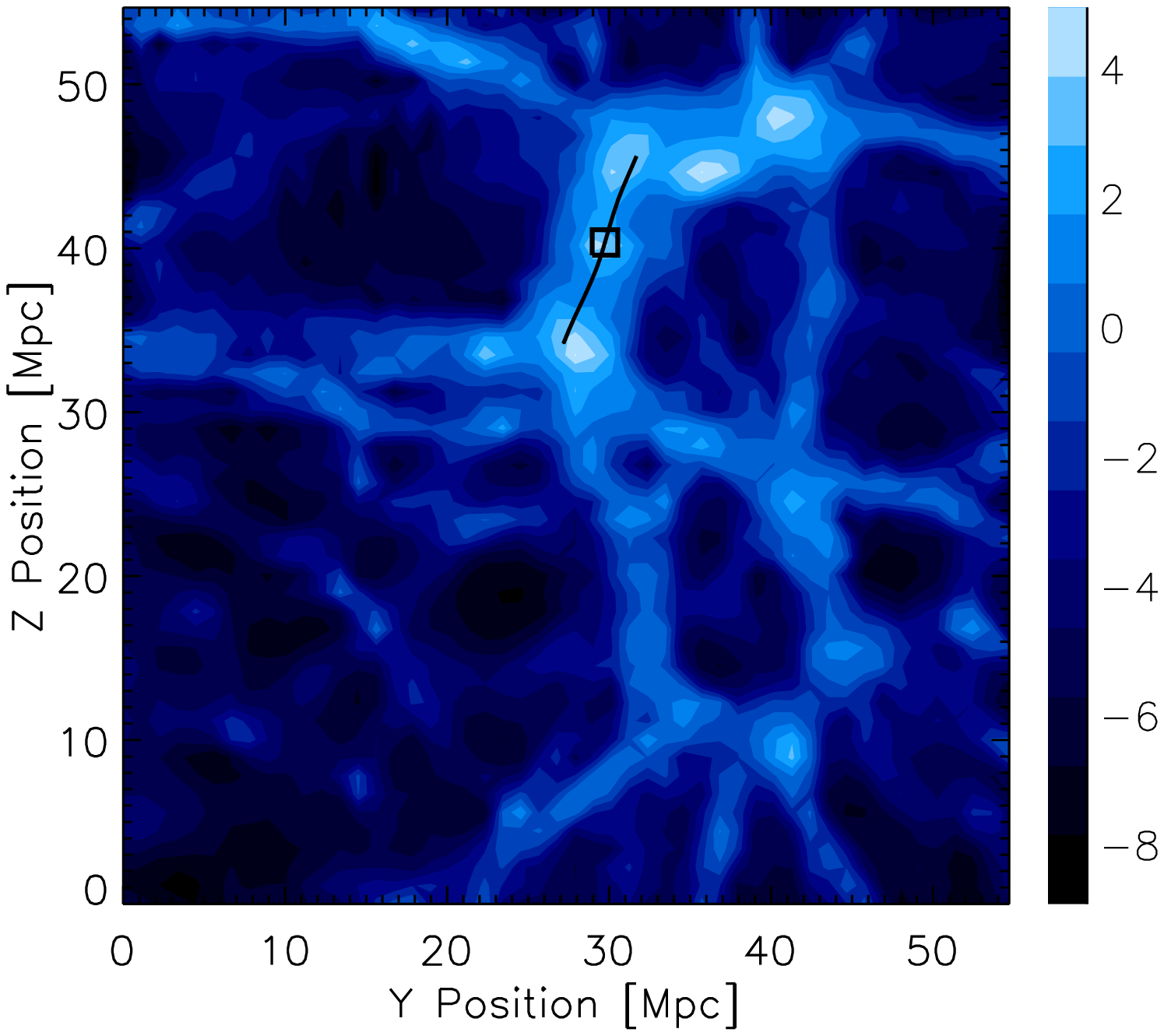}
   \caption{Distribution of the magnetic field intensity in the region
     surrounding an ultrahigh energy cosmic ray source. Each panel is
     a 1.1~Mpc thick (size of a grid cell) slice of the simulated
     Universe, cut along the axis perpendicular to the observed
     plan. From left to right, the distribution is of type
     ``isotropic" (Eq.~\ref{eq:iso}), ``anisotropic"
     (Eq.~\ref{eq:ani}) and ``contrasted" (Eq.~\ref{eq:contrast}). The
     color code is the same for all the panels and the values
     indicated on the right hand side correspond to the logarithm of
     the magnetic field intensity in nano Gauss. The black line
     indicates the axis of the filament that is studied throughout
     this paper and the black square the chosen position of the
     source. The normalization factor equals $B_0=1$~nG.}
     \label{fig:slices}
\end{figure*}

\subsection{Magnetic field configuration}

Our current knowledge on the structure of magnetic fields at large
scales is very poor. It mainly stems from the lack of observations due
to the intrinsic weakness of the fields. Moreover, no satisfactory
theory has been established to explain the origins of these fields and
the further magnetic enrichment of the Universe. There is now a decade
of history of numerical simulations that endeavor to model the
distribution of the magnetic field at large scales. One may cite in
particular the works of \cite{Ryu98}, \cite{DGST04,DGST05},
\cite{SME04} and most recently \cite{Das08}. In these studies,
magnetized seeds are injected in a cosmological simulation and the
overall magnetic field then evolves coupled with the underlying
baryonic matter. The final amplitude of the field is set to adjust the
measured intensity inside clusters of galaxies (in \citealp{Das08} the
amplitude is not arbitrary but is directly estimated from the gas
kinetic property and corresponds nevertheless to the observations
inside clusters). Unfortunately, these various simulations do not
  converge as to the present-day configuration of extragalactic
  magnetic fields (see for example Fig.~1 of \citealp{KL08a}) and
  consequently, they also diverge as to the effect of these magnetic
  fields on the transport of ultrahigh energy cosmic rays. The cause
  of this difference likely lies in the choice of initial data, which
  cannot be constrained with our current knowledge. Given the fact
  that such simulations are time consuming, these tools cannot be
  considered as practical.

In view of this situation, we explore the influence of three typical
types of magnetic field structures on the gamma ray emission,
following the method developed by \cite{KL08a}. We map the magnetic
field intensity distribution according to the underlying matter
density $\rho$, assuming the relation $B =B_0\,f(\rho)$, where $B_0$
is a normalization factor, and the dimensionless function $f$ may be
modelled as:
\begin{eqnarray}
&f_{\rm iso}(\rho)&\,=\,\rho^{2/3}\ ,\label{eq:iso} \\
&f_{\rm ani}(\rho)&\,=\,\rho^{0.9}\ , \label{eq:ani}\\ 
&f_{\rm contrast}(\rho)&\,=\,\rho\left[1+\left({\rho\over\langle\rho\rangle}\right)^{-2}\right]^{-1}.\label{eq:contrast}
\end{eqnarray} 
Note that $B_0$ corresponds approximately to the mean value of the
magnetic field in the Universe.  These phenomenological relationships
greatly simplify the numerical procedure to obtain a map of the
extragalactic magnetic field since a pure dark matter (i.e. non
hydrodynamical) simulation of large scale structure provides a
sufficiently good description of the density field. These
relationships are motivated by physical arguments related to the
mechanism of amplification of magnetic fields during structure
formation and they encapture the scalings observed in numerical
simulations up to the uncertainty inherent in such simulations (see
the corresponding discussion in \citealp{Dolag06} and
\citealp{KL08a}).  The first relation is typically expected if
baryonic matter undergoes an isotropic collapse.  In dense regions of
the Universe where viscosity and shear effects govern the evolution of
the magnetized plasma however, the relation would get closer to
Eq.~(\ref{eq:ani}). This formula can be derived analytically in the
case of an anisotropic collapse along one or two dimensions, in
filaments or sheets for example \citep{KC06}.  The last model is an
ad-hoc modeling of the suppression of magnetic fields in the voids of
large structure which leaves unchanged the distribution in the dense
intergalactic medium (meaning $\rho>\langle\rho\rangle$). It thus
allows to model a situation in which the magnetic enrichment of the
intergalactic medium is related to structure formation, for instance
through the pollution by starburst galaxies and/or radio-galaxies (see
the corresponding discussion in \citealp{KL08b}). In the following, we
will refer to these three models as ``isotropic", ``anisotropic" and
``contrasted" respectively.

In order to have a representative sample of the underlying matter
density at large scales, we use a three-dimensional output (at
redshift $z=0$) of a cosmological Dark Matter simulation provided by
S.~Colombi. It assumes a $\Lambda$CDM model with
$\Omega_\mathrm{m}=0.3$, $\Omega_\mathrm{\Lambda}=0.7$ and Hubble
constant $h \equiv H_0/(100$ km s$^{-1}$ Mpc$^{-1}) = 0.7$. The
simulation models a $200\, h^{-1}$~Mpc comoving periodic cube split in
$256^3$ cells, where the Dark Matter overdensity is computed. We do
not resolve structures below the Jeans length, which implies that we
can identify the computed Dark Matter distribution to a gas
distribution.

\subsection{Numerical code}\label{subsubsection:method}

We compute the gamma ray signal through numerical Monte Carlo
simulations, using the code presented in \cite{KAM09}. In a first
step, we propagate ultrahigh energy cosmic rays (protons and heavier
nuclei) in one of the three-dimensional magnetic field models
described above. The coherence length $\lambda_B$ of the magnetic
field is assumed to be constant throughout space and is set to
$100\,$~kpc. This certainly is somewhat ad-hoc but: the coherence
  scale enters the calculation in the deflection angle under the form
  $B\sqrt{\lambda_B}$, hence discussing various normalizations of the
  magnetic field strength, as we do further below, allows to encapture
  different possible values for $\lambda_B$; furthermore, $100\,$kpc
  seems a reasonable compromise between an upper limit of a few
  hundreds of kpc for $\lambda_B$ based on the turn-around time of
  largest eddies ~\citep{Waxman:1998yy} and the typical galactic scales
  $\sim\,10-30\,$kpc, see \cite{KL08a} for a detailed discussion. The
generation of secondary photons and electrons and positrons through
pion production is treated in a discrete manner, as in \cite{Allard06}
for nuclei projectiles with $A> 1$ and with SOPHIA \citep{Mucke99} for
proton and neutron cosmic rays. Electron and positron pair creation
through photo-hadronic processes is implemented as in
\cite{ASM06}. For each time step (chosen to be much larger than the
typical mean free path of pair photo-production processes), we assume
that an ensemble of electron and positron pairs are generated with
energies distributed according to a power law, and maximum energy
depending on the primary particle. The use of an improved pair
spectrum as computed in \cite{KA08} would lower the gamma ray flux
resulting from direct pair production by a factor of a few in the
range $E_{\gamma}\lesssim 10\,$GeV, and leave unchanged the prediction
at a higher energy (Armengaud, private communication). Furthermore, we
will show in the following that direct pair production processes
provide a sub-dominant contribution to the overall gamma ray flux in
the range $E_{\gamma} \,\gtrsim\,0.1\,$GeV, hence the overall
difference is expected to be even smaller.

We investigate in this paper the possible detection of photons in the
GeV-TeV energy range. Equation~(\ref{eq:Esyn}) suggests that, unless
the source is embedded in a particularly strong field, the major
contribution in this range will come from electrons of energy
$E_e\gtrsim 10^{18}$~eV. The flux will thus essentially result from
cosmic rays of energy higher than $E\sim 10^{19}~$eV. For this reason,
we chose to inject cosmic rays at the source between $E_{\rm
  min}=10^{17}$~eV et $E_{\rm max}=10^{20.5}$~eV. We assume our source
to be stationary with (isotropic) cosmic ray luminosity integrated over
$E=10^{19}$~eV of $L_{E,19} = 10^{42-46}~$erg~s$^{-1}$, and a spectral
index of $\alpha = 2.3$.

We take into account photo-hadronic interactions with CMB and infrared
photons. The diffuse infrared background is modelled according to the
studies of \cite{SMS06}. We do not take into account redshift
evolution, as its effect is negligible as compared to the
uncertainties on our other parameters. We also neglect baryonic
interactions in view of the negligible density in the source
environment.

Once the secondary particles produced during the propagation have been
computed, we calculate the synchrotron photon fluxes, taking into
account the competition with inverse Compton scattering by Monte Carlo
over steps of size 100~kpc. The mean effective electron loss length is
calculated following \cite{GA05}, using the radio background data of
\cite{CBA70}. We assume that each electron emits a synchrotron photon
spectrum ${\rm d}N_\gamma/{\rm d}E_\gamma \propto E_\gamma^{-2/3}$
with a sharp cut-off above $E_{\rm c}\sim (2/3)\,E_{\rm \gamma,syn}$,
where $E_{\rm \gamma,syn}$ is given by Eq.~(\ref{eq:Esyn}).

Regarding the photons produced through the neutral pion channel, we
draw the interaction positions with CMB and radio photons using the
mean free paths computed by \cite{Lee98}. We assume that one of the
electron-positron pair inherits of the total energy of the parent
photon.

\begin{figure*}[tbhp]
\begin{center}
\includegraphics[width =\textwidth]{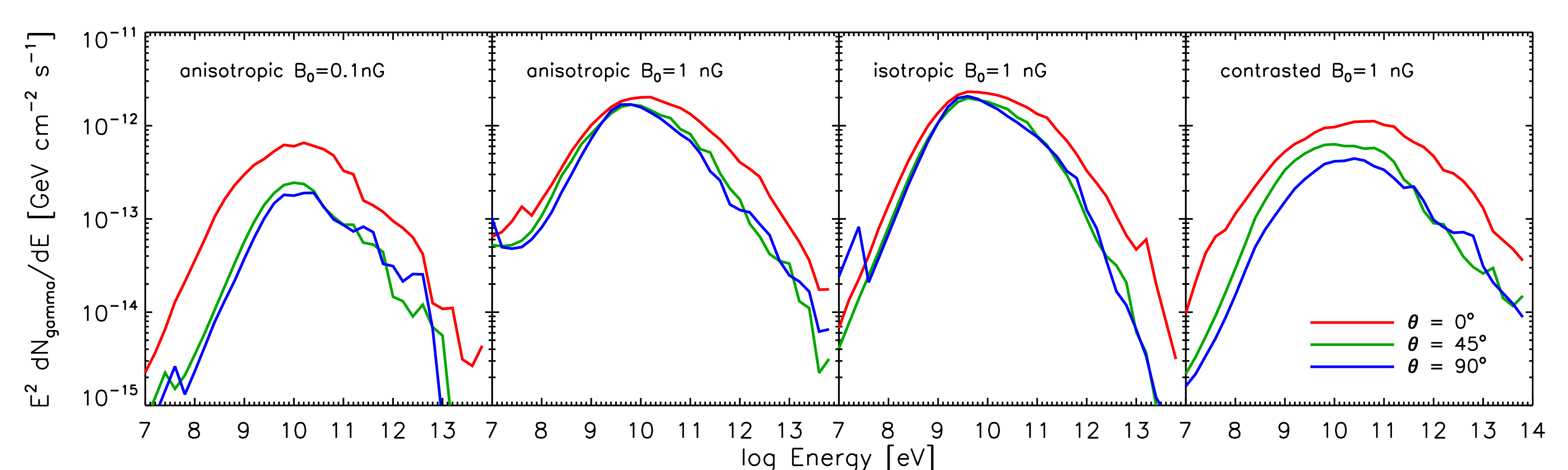}
\caption{Flux of photons produced by synchrotron emission for a
  filament observed from the directions forming the indicated angles
  with its axis. The filament is located at 100~Mpc and harbors a
  source that injects a pure proton composition with (isotropic) luminosity of
  $L_{E,19}=10^{42}$~erg/s and a spectral index of 2.3.  The four
  panels represent various models of magnetic fields. From left to
  right: ``anisotropic'' with $B_0=0.1~$nG, ``anisotropic" with
  $B_0=1~$nG, ``isotropic" with $B_0=1~$nG and ``contrasted" with
  $B_0=1~$nG.}  \label{fig:angles_fil}
\end{center}
\end{figure*}

\subsection{Synchrotron signal for various magnetic field configurations}\label{subsubsection:res_mag}

We first study the influence of inhomogeneous magnetic fields on the
gamma ray flux produced by ultrahigh energy cosmic rays near their
source. In particular, we examine the case of a source placed in a
rather dense region of a filament of large scale structure, see
Fig.~\ref{fig:slices}. On the transverse scale of the structure
$l_\perp$, of order of a few Mpc, one expects the secondary electrons
and positrons to be distributed isotropically around the source. A
population of ultrahigh energy cosmic rays seeds the intergalactic
medium with secondary pairs in a roughly homogeneous manner on the
energy loss length scale of the interaction process $\lambda$,
i.e. the number of electrons/positrons injected per unit time per unit
distance to the source ${\rm d}N_e/{\rm d}r{\rm d}t$ does not depend
on $r$ as long as $r\ll\lambda$.  For ultrahigh energy protons, this
length scale is $\sim1\,$Gpc above $10^{19}\,$eV for pair production
and $\sim 100\,$Mpc above $10^{20}\,$eV for pion production. From the
point of view of secondary synchrotron $\gamma$ emission, however, the
emission region is limited to the fraction of space in which the
magnetic field is sufficiently intense for synchrotron cooling to
predominate over inverse Compton cooling.
  
The bulk of gamma rays is produced by electrons of energy $E_{\rm
  e}\gtrsim 10^{19}~$eV, themselves produced through pion production
by protons of $E_{\rm p}\gtrsim 10^{20}~$eV.  In the absence of
significant deflection experienced by these primary particles, one
expects the synchrotron emission to be enhanced along the filament
axis, where the magnetic field keeps a high value over a longer length
than along the perpendicular direction (see
Fig.~\ref{fig:slices}). The whole structure where
$x_{eB}<x_{e\gamma}$, i.e. $B_{\rm IGM} \gtrsim 0.5\,\mbox{nG} \,
(E_e/10^{19}\,{\rm eV})^{-3/4}$ is then illuminated in synchrotron
from electrons and positrons provided $x_{eB}< c t_{\rm esc}$, where
$t_{\rm esc}$ denotes the escape timescale from the magnetized
region. For a magnetic field with $B\sim 10\,$nG and coherence length
$\lambda_B\lesssim 100\,$kpc, the Larmor time $t_{\rm L}\,\sim\,1\,{\rm
  Mpc}\,(E_e/10^{19}\,{\rm eV})(B/10\,{\rm nG})^{-1}\,\sim\lambda_B$,
hence confinement is marginal and the escape length $t_{\rm esc}\sim
l_\perp \gg x_{eB}$ (see \citealp{CLP02} for more details on the
escape length).

At a given energy $E_e$, the ratio between the fluxes along and
perpendicular to the filament should be of order $a/b$, where $a$ and
$b$ are the characteristic lengths of the axis of the filament where
$B_{\rm IGM} \gtrsim 0.5\,\mbox{nG} \, E_{e,19}^{-3/4}$.
  
Figure~\ref{fig:angles_fil} depicts indeed such an effect. It
represents the synchrotron emission produced by a filament at a
distance of 100~Mpc, embedding a source of luminosity
$L_{E,19}=10^{42}$~erg~s$^{-1}$ and an injection spectral index of
2.3.  One can notice that the difference between the highest and
lowest flux values is of a factor $\sim 2-5$, which corresponds to the
ratio between the filament major axis. For a narrower or longer
filament, this interval would be more pronounced -- which does not
necessarily mean that the flux would be enhanced along the filament
axis.

This difference of flux according to the observation angle disappears
at low energy for the second and third panels of
Fig.~\ref{fig:angles_fil}. This is intricately related to the combined
effect of the configuration of the magnetic field and the ratio of
electrons producing the observed peak of gamma rays. It corresponds to
a coincidence and should not be viewed as an effect of diffusion of
primary protons. We tested indeed that these differences remain present when cosmic rays are forced to propagate rectilinearly. For the contrasted magnetic field, the gamma-ray fluxes are overall lower, as the field strength diminishes sharply outside of the central part of the filament. The orthogonal section of the filament is thus substantially magnetized only over a short length, which consequently reduces the gamma-ray flux observed from $90^\circ$. For the isotropic and anisotropic cases with $B_0=1~$nG, electrons and positrons experience a weaker field on average when they cross the filament orthogonally, instead of traveling through their entire length. For the flux observed at $90^\circ$, photons with energy $<10~$GeV are thus generated by electrons of slightly higher energy ($E_{90}$) than in the $0^\circ$ case ($E_0$). By coincidence, $E_{90}$ corresponds approximately to the peak of distribution of secondary electrons, whereas $E_0$ is shifted according to the peak. The flux observed at $90^\circ$ is thus amplified for energies $<10$~GeV due to the enhanced number of electrons contributing to the emission, which brings up the fluxes at various observation angles closer. Finally, for the anisotropic case with $B_0=0.1$~nG, the emission is suppressed when the filament is not observed along its major axis, because of the weak mean field that particles experience (the mean field remains reasonably strong along the axis). The gap between the flux at $0^\circ$ and the other cases stems from this effect.

Most of all, Fig.~\ref{fig:angles_fil} demonstrates that the photon
flux around $E_\gamma\sim10~$GeV is fairly robust with respect to the
various magnetic field configurations, which can be understood from
Fig.~\ref{fig:slices}: the extent of the magnetized region above
$B\sim 1~$nG is indeed not fundamentally modified by the chosen
distribution and the photon flux is hardly more sensitive to the
overall intensity of the field (in particular, we tested that a value
of $B_0=10$~nG does not affect the results of more than a factor 2,
for all magnetic configurations). All in all, for a quite reasonable
normalization of the magnetic field, meaning an average
$B_0\sim0.1-10\,$nG at density contrast unity, and for various
  scalings of the magnetic field with density profile, the variations
in flux remain smaller than an order of magnitude in the range
$E_\gamma \sim 10-1000\,$GeV.

Figure~\ref{fig:angles_fil} also illustrates that the energy at
  which the signal peaks depends only weakly on the normalization of
  the magnetic field $B_0$, and this remains true up to a few
  nano-Gauss. This can be understood qualitatively, assuming for
  simplicity the field to be homogeneous over the area of interest,
  with strength $B_0$. For $B_0=1\,$nG, the electron cross-over energy
  $E_\times$ beyond which electrons radiate in synchrotron reads
  $E_\times\simeq 10^{18}\,$eV (see the discussion in
  Sec.~\ref{section:B}). However, the energy spectrum $E_e {\rm
    d}N_e/{\rm d}E_e$ of secondary electrons deposited in the source
  vicinity peaks at $\sim10^{19}\,$eV, as a result of the competition
  between the opposite scalings with energy of the abundance of parent
  protons and of the parent proton energy loss rate. Therefore, the
  synchrotron signal is expected to peak at $E_\gamma \sim 10\,{\rm
    GeV}$, see Eq.~\ref{eq:Esyn} with $B_0=1\,$nG and
  $E_e=10^{19}\,$eV. Now, for $B_0=0.1\,$nG, the cross-over energy
  $E_\times \simeq 4\times 10^{19}\,$eV so that one should use
  $E_e=4\times 10^{19}\,$eV in Eq.~\ref{eq:Esyn} with $B_0=0.1\,$nG,
  giving $E_\gamma \sim 10\,$GeV again: the larger typical electron
  energy (among those radiating in synchrotron) compensates for the
  smaller value of $B_0$. As $B_0$ is increased significantly above
  $1\,$nG, the peak location of the gamma-ray signal will increase in
  proportion, as most of the secondary electron flux can be
  radiated in synchrotron.

Note that the flux should be cut off above $E_{\gamma}\sim 10$~TeV due
to the opacity of the Universe to photons above this energy. This
effect is not represented in these plots for simplicity, as its
precise spectral shape depends on the distance $d$ to the source
(while the sub-TeV gamma ray flux scales in proportion to $L_{\rm
  cr}/d^2$).

At lower energy ($E_{\gamma}\lesssim 1$~GeV), the flux intensity can
vary by more than two orders of magnitude depending on the chosen
magnetic configuration. In the low energy range, magnetic confinement
indeed starts to play an important role. Furthermore, the inverse
Compton energy loss length shortens drastically at these energies so
that only strong magnetic fields can help avoiding the formation of an
electromagnetic cascade and lead instead to synchrotron emission in
the filament.

\subsection{Synchrotron signal for various chemical compositions and injection spectra}\label{section:compo}

In this section, we discuss the effect of the injected chemical
composition and of the spectral index at the source. As discussed
above, there are conflicting claims as to the measured chemical
composition at ultrahigh energies; in particular, the HiRes experiment
reports a light composition \citep{Hires10} while the Pierre Auger
Observatory measurements point toward a composition that becomes
increasingly heavier at energies above the ankle
\citep{Auger10}. Theory is here of little help, as the source of
ultrahigh energy cosmic rays is unknown; protons are usually
considered as prime candidates because of their large cosmic
abundance, but at the same time one may argue that a large atomic
number facilitates acceleration to high energy.
  
In this framework, \cite{LW09} have proposed a test of the chemical
composition of ultra-high energy cosmic rays on the sky, using the
anisotropy patterns measured at various energies, instead of relying
on measurements of the depth of maximum shower development. It is
shown in particular that, at equal magnetic rigidities $E/Z$, one
should observe a comparable or stronger anisotropy signal from the
proton component than from the heavy nuclei component emitted by the
source, even if the source injects protons and heavy nuclei in equal
numbers at a given energy. When compared to the 99\% c.l. anisotropy
signal reported by the Pierre Auger Observatory at energies above
$5.7\times10^{19}\,$eV \citep{Auger1,Auger2}, one concludes that: if the composition is
heavy at these energies, and if the source injects protons in at least
equal number (at a given energy) as heavy nuclei, then one should
observe a comparable or stronger anisotropy pattern above
$5.7\times10^{19}/Z\,$eV. 

Future data will hopefully cast light on this issue, but in the
meantime it is necessary to consider a large set of possible chemical
compositions. We thus consider the following injection spectra:
\begin{enumerate}
\item Two pure proton compositions with indices $\alpha = 2.3$ and
  2.7. A soft spectral index of $\sim 2.7$ seems to be favored to fit
  the observed cosmic ray spectrum in the case of a pure proton
  composition, especially in the ``dip-model" proposed by
  \cite{BGG06}, where the transition between the Galactic and
  extragalactic components happens at relatively low energy ($E\sim
  10^{17.5}$~eV).
\item A proton dominated mix composition with spectral index
  $\alpha=2.3$, based on Galactic cosmic ray abundances as in
  \cite{Allard06}. Such a spectrum is mostly proton dominated at
    ultra-high energies.
\item A pure iron composition with spectral index $\alpha=2.3$.
\item A mixed composition that was proposed by \cite{Allard08}, that
  contains 30\% of iron. In this injection, the maximum proton energy
  is $E_{{\rm max},p}=10^{19}~$eV (\citealp{Allard08} considered
  $E_{{\rm max},p}=4\times10^{19}~$eV but an equally satisfying fit ot
  the spectrum can be obtained for a lower value). Assuming that heavy
  elements are accelerated at an energy $Z\times E_{{\rm max},p}$, and
  because of the propagation effects that disintegrate intermediate
  nuclei preferentially as compared to heavier nuclei, one would then
  measure on the Earth a heavy composition at the highest energy. The
  resulting detected spectrum allows to reproduce the maximum depth of
  shower measurements of the Pierre Auger Observatory. However, the
  anisotropy pattern should also appear in this case at energies of
  order a few EeV, as discussed above.  An injection spectrum of 2.0
  is necessary to fit the observed cosmic ray spectrum.
\end{enumerate}
For the first three cases, the maximum proton injection energy is 
$E_{{\rm max},p}=10^{20.5}~$eV. In all cases, we assume $E_{{\rm
    max},Z}=Z\times E_{{\rm max},p}$ for a nucleus of charge number
$Z$ (an exponential cut-off is assumed).

\begin{figure}[tbhp]
\centering
\resizebox{\hsize}{!}{\includegraphics{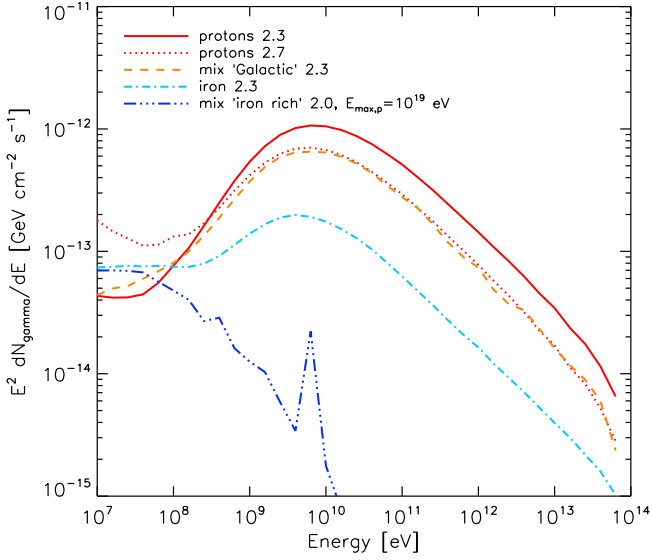}}
\caption{ Photon flux produced by synchrotron emission for different
  source injections. We chose an ``anisotropic" distribution of the magnetic field with normalization $B_0 = 1$~nG. The source has a luminosity of
  $L_{E,19}=10^{42}$~erg~s$^{-1}$, and is located in the filament
  described previously, at a distance of 100~Mpc. The average flux
  integrated over all angular lines of sight are
  presented. }  \label{fig:gamma_compo}
\end{figure}

Figure~\ref{fig:gamma_compo} presents the fluxes obtained for these
different injection spectra. The source is again located in the
filament described earlier, at a distance of 100~Mpc, and has a
luminosity above $E=10^{19}$~eV of
$L_{E,19}=10^{42}$~erg~s$^{-1}$. Due to the comparatively larger
number of protons at low energy, the photon flux below 1~GeV is
amplified in the case of an injection spectrum of 2.7 as compared to a
2.3 index. For similar reasons, the flux at higher energies ($\gtrsim
10\,$GeV) is lower.

The shift in amplitude between the fluxes produced by the different
compositions can be understood as follows. The energy of the electrons
and positrons produced by photo-disintegration is proportional to the
nucleus Lorentz factor $E_e\propto \Gamma_A \propto E_A/A$. Hence at a
given electron energy, the amount of injected energy by heavy nuclei is 
reduced if the spectral index is $\ge 2$. Besides, pion
production via baryonic resonance processes on the CMB photons happens
at energies higher than for protons ($E\gtrsim10^{20}$~eV). Obviously,
the flux produced by the mix ``Galactic" composition with a 2.3
injection differs only slightly from that produced by protons with the
same spectral index as such a composition is dominated by
protons.

\begin{figure}[tbhp]
\centering
\resizebox{\hsize}{!}{\includegraphics{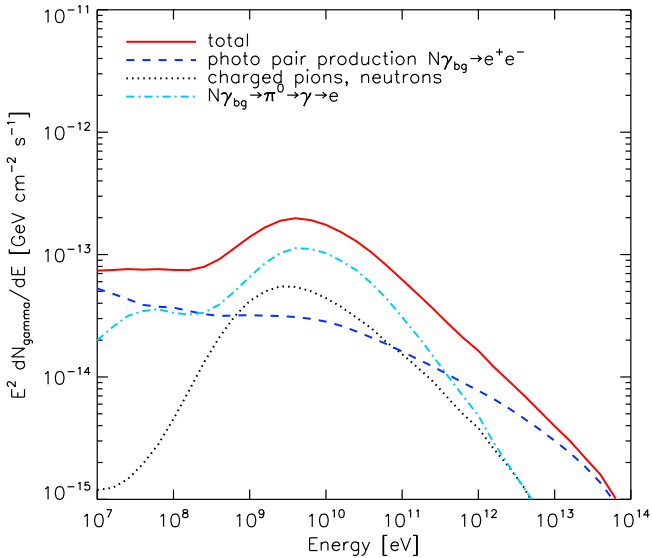}}
\caption{Same as Fig.~\ref{fig:gamma_compo}, but for the particular
  case of a pure iron injection with spectral index 2.3. In red solid
  lines, the total flux, in blue dashed lines the contribution of the
  photo-pair production, in black dotted lines the flux due to charged
  pion production via photo-disintegration processes and to neutron
  disintegration, in pale blue dash-dotted lines, the contribution of
  the neutral pion channel. }  \label{fig:gamma_fer}
\end{figure}

The above trends can also be noted in Fig.~\ref{fig:gamma_fer}, which
reveals a smaller contribution of photo-disintegration processes (in
black dotted and pale blue dash-dotted lines) than for a proton
composition. The yield of the electrons and positrons produced through
photo-pair production (in blue dashed lines) is not as strongly
attenuated however, as the secondary nucleon are emitted above the
pair production threshold (but below the pion photo-production
threshold). The presence of a second bump at lower energy in the
neutral pion channel (pale blue dash-dotted) is related to the
production of photons via the giant dipole resonance process.

Finally, the flux is suppressed in the case of an iron enriched mix
composition, because of the relatively low cut-off energy, $E_{{\rm
    max},p}=10^{19}\,$eV. In this case, $E_{{\rm max},p}$ is
  lower than the pion production threshold energy for protons and the
  heavy nuclei maximum energy is also lower than the pion production
  threshold, so that this channel of secondary $e^+e^-$ injection
  disappears.

\section{Discussion on detectability}\label{section:detectability}

\subsection{Synchrotron signal from equal luminosity sources}\label{subsection:average}

Our study indicates that around $\sim 10-100~$GeV, the gamma ray flux
produced by the propagation of ultrahigh energy cosmic rays is fairly
robust to changes in composition and in the configuration of magnetic
fields. The normalization of the flux ultimately depends on the
injected energy, namely, the cosmic ray luminosity and the injection
spectral index at the source. The cosmic ray luminosity of the source
may be inferred through the normalization to the measured flux of
ultrahigh cosmic rays once the source density is specified. The
apparent density of ultrahigh energy cosmic ray sources, which
coincides with the true source density $n_{\rm s}$ for steady isotropic emitting
sources (which we assume here; see Section~4 for a brief discussion of
bursting/beamed sources), is bound by the lack of repeaters in the Pierre
Auger dataset: $n_{\rm s}\gtrsim
10^{-5}\,$Mpc$^{-3}$~\citep{KW08}. This lower bound may
in turn be translated into an upper limit for the gamma ray flux
expected from equal luminosity cosmic ray sources, $L_{E,19}\lesssim
10^{42}\,$erg/s.

\begin{figure}[tbh]
\begin{center}
\includegraphics[width=0.5\textwidth]{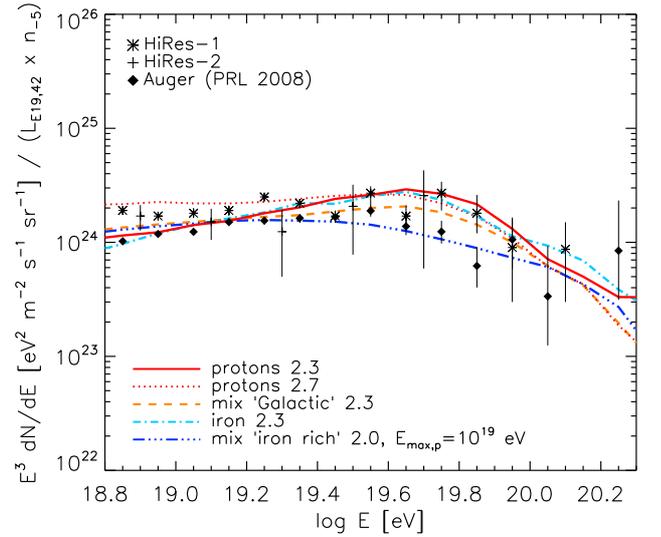} 
\caption{Propagated spectra of ultrahigh energy cosmic rays for
  various chemical compositions as described in
  section~\ref{section:compo} and corresponding spectral indices to
  best fit the observational data. Sources are assumed to share an
  equal luminosity $L_{E,19}=10^{42}$~erg$\,$s$^{-1}$ and a spatial
  density $n_{\rm s}=10^{-5}$~Mpc$^{-3}$.}  \label{fig:spectra_fil}
\end{center}
\end{figure}

Figure~\ref{fig:spectra_fil} presents the spectra obtained for
sources of luminosity $L_{E,19}=10^{42}$~erg$\,$s$^{-1}$ and of number
density $n_{\rm s}=10^{-5}$~Mpc$^{-3}$. For each case, we represent
the median spectrum of 1000 realizations of local source
distributions, the minimum distance to a source being of 4~Mpc. Our
calculated values are comparable to the cosmic ray fluxes detected by
the two most recent experiments: HiRes and the Pierre Auger
Observatory. The gamma ray fluxes obtained for single sources with
this normalization (i.e. a luminosity of
$L_{E,19}=10^{42}$~erg$\,$s$^{-1}$) at a distance of 100~Mpc are
presented in Figs.~\ref{fig:angles_fil}, \ref{fig:gamma_compo}
and~\ref{fig:gamma_fer}.

These gamma ray fluxes lie below the limits of detectability of
current and upcoming instruments such as Fermi, HESS or the Cerenkov
Telescope Array (CTA), by more than two orders of magnitude. The
sensitivity of the Fermi telescope is indeed of order $\sim
2\times10^{-10}~$GeV s$^{-1}$ cm$^{-2}$ around 10~GeV for a year of
observation and the future Cherenkov Telescope Array (CTA) is expected
to have a sensitivity of order $\sim 10^{-11}$~GeV~$\rm s^{-1}
cm^{-2}$ for 100 hours for a point source around TeV energies \citep{Atwood09,Wagner09}.

\subsection{Rare and  powerful sources}

We conclude from the previous subsection that only powerful sources
embedded in magnetized environments, rare with respect to the average
population of ultrahigh energy cosmic ray sources, may produce a gamma
ray signature strong enough to be observed by current and upcoming
experiments. In the following, we examine the case of two such sources
embedded in filaments, with (isotropic) luminosity chosen to be higher than the
average allowed by the cosmic ray normalization for equal luminosity
sources: $L_{E,19}=10^{44}$~erg~s$^{-1}$ and
$L_{E,19}=10^{46}$~erg~s$^{-1}$. As a case study, we place these
sources at a distance of $D=100~$Mpc and $D=1~$Gpc respectively. Those
distances are consistent with the number density of objects found at
comparable photon luminosities (see for example \citealp{Wall05}).

\begin{figure}[tbh]
\begin{center}
\includegraphics[width=0.5\textwidth]{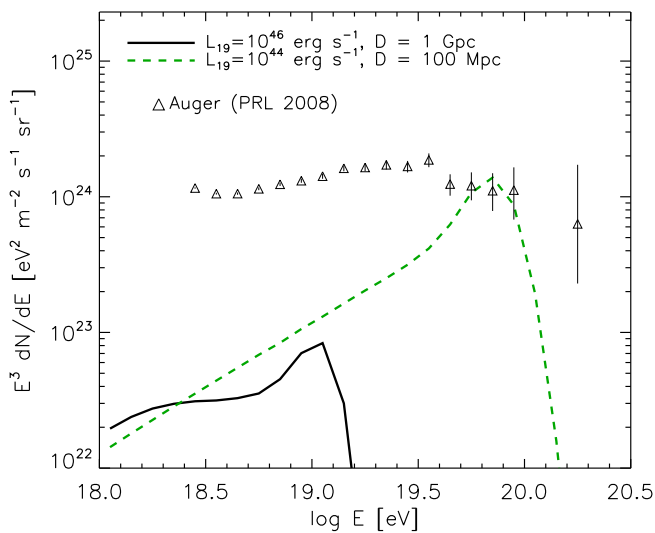} 
\caption{Propagated spectra of ultrahigh energy cosmic rays from a
  single source with luminosity $L_{E,19}=10^{46}$~erg~s$^{-1}$
  located at a distance of $D=1~$Gpc (black solid) or
  $L_{E,19}=10^{44}$~erg~s$^{-1}$ and $D=100~$Mpc (green dashed). The
  injection spectral index is 2.0 in both cases. The observed Auger
  spectrum \citep{Auger08} is overlaid.} \label{fig:onesource}
\end{center}
\end{figure}

The ultrahigh energy cosmic ray spectra obtained after propagation
from these two model sources are represented in
Fig.~\ref{fig:onesource}. The injection at the source is assumed to be
pure proton and with an index of 2.0. The comparison with the Auger
data shows that the close-by source (green dashed) is marginally
excluded. More precisely, such a source should produce a strong
anisotropy signal, of order unity relatively to the all-sky
background, if the magnetic deflection at this energy $\sim 7\times
10^{19}\,$eV is smaller than unity. Following \cite{KL08b}, the
expected deflection is of order $2^\circ Z$ if one models the
magnetized Universe as a collection of magnetized filaments with
$B=10\,$nG ($\lambda_B=100\,$kpc) immersed in unmagnetized voids with
typical separation $\sim40\,$Mpc. This estimate neglects the
deflection due to the Galactic magnetic field, which depends on
incoming direction and charge. Qualitatively speaking, however, one
would then expect the above source to produce a strong anisotropy
signal unless it injects only iron group nuclei with $Z\sim26$ at the
highest energies. In this latter case, however, the gamma-ray signal
would be lower by a factor $\sim 5$ or even suppressed if $E_{\rm
  max,p}$ is less than the pion production threshold, as discussed in
the previous Section. Furthermore, such a source might also give rise
to a strong anisotropy at energies $\sim 3\,$EeV, depending on the
composition ratio of injected protons to iron group nuclei
\citep{LW09}.

On the contrary, the remote source (indicated by the black solid line
in Fig.~\ref{fig:onesource}) contributes to $\sim 10\%$ of the total
flux around $E_{\rm cr}\sim 10^{19}$~eV. Such a source would be wholly
invisible in ultrahigh energy cosmic rays, because of the large
expected magnetic deflection at this energy, on a large path length of
$1\,$Gpc.

\begin{figure*}[tbhp]
\centering
\includegraphics[width =0.47\textwidth]{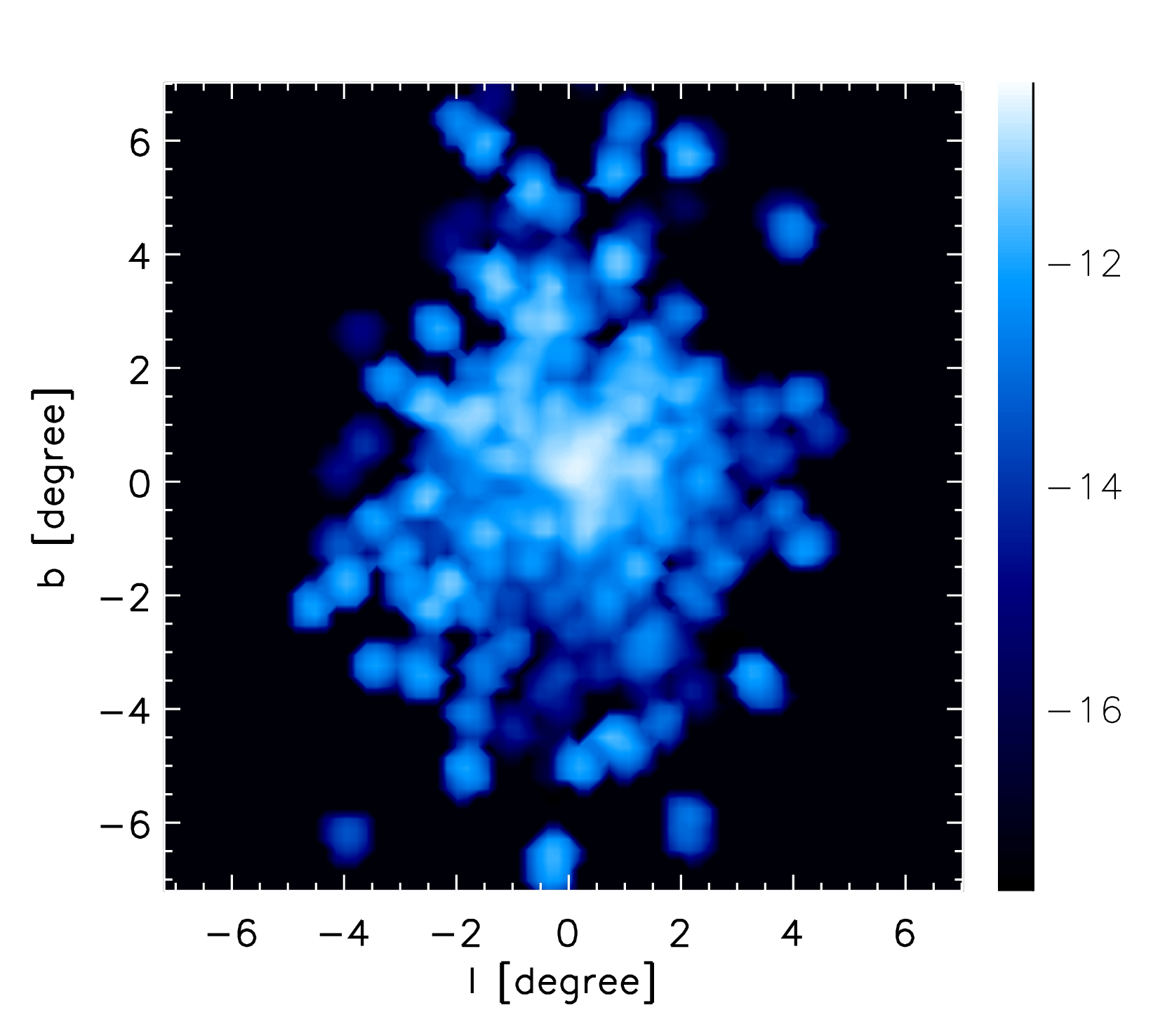}
\includegraphics[width =0.47\textwidth]{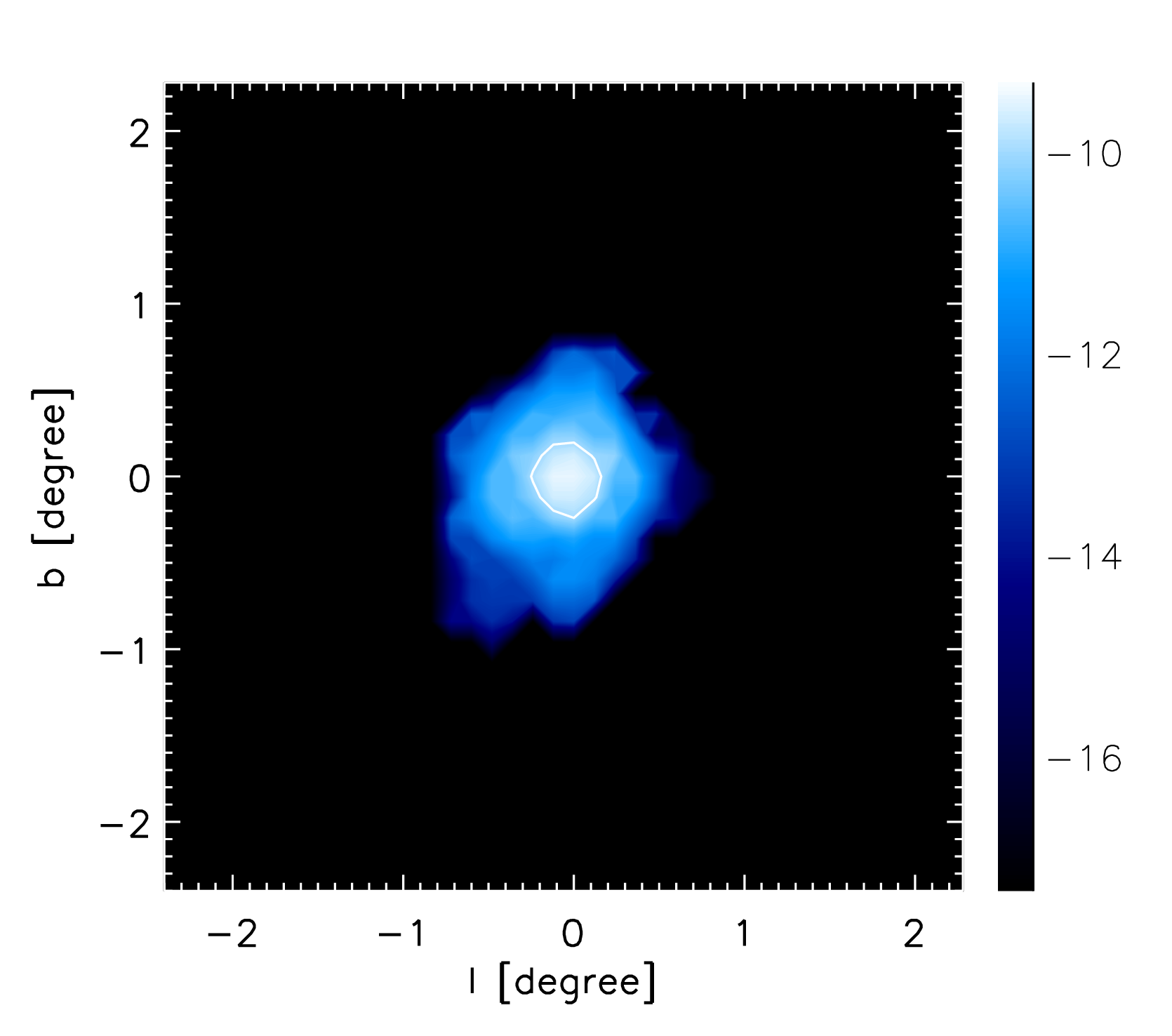}
\caption{Image of a filament in gamma rays produced via synchrotron
  emission, integrated over $E_\gamma=1-100~$GeV. We chose an ``anisotropic" distribution of the magnetic field with normalization $B_0 = 1$~nG. {\it Left}: the
  filament is observed along its axis from a distance of 100~Mpc and
  the embedded source has a cosmic ray luminosity of
  $L_{E,19}=10^{44}$~erg~s$^{-1}$. {\it Right}: the filament is
  observed along its axis from a distance of 1~Gpc and the embedded
  source has a luminosity of $L_{E,19}=10^{46}$~erg~s$^{-1}$. The
  color bar indicates the logarithm of the flux in
  GeV~cm$^{-2}$~s$^{-1}$. White contours indicate a flux level of
  $10^{-10}~$GeV~cm$^{-2}$~s$^{-1}$.}  \label{fig:filim}
\end{figure*}

Let us now discuss the possibility of detecting these sources in gamma
rays. Figure~\ref{fig:filim} presents our computed images of
synchrotron gamma ray emission for these two examples of sources
embedded in the filament shown in Fig.~\ref{fig:slices}.  

However, one should be aware that the sensitivity of gamma ray
telescopes weakens for extended sources by the ratio of the radius of
the emission to the angular resolution. Typically, Fermi LAT and HESS
have an angular resolution of some fractions of a degree around
energies of 10~GeV and 100~GeV respectively. CTA will have an even
higher resolution of the order of the arc-minute above 100~GeV. For
this reason, one will have to find a source with an acceptable range
in luminosity and distance, so as to not lose too much flux due to the
angular extension, but still being able to resolve this extension. Let
us note that the magnetic configuration around the source should have
a negligible role on the flux intensity, as we demonstrated in
section~\ref{section:B}. However, the magnetic configuration, in
particular the strength and the coherence length, directly controls
the extension of the image, which is given by the transverse
displacement of the protons through magnetic deflection~\citep{GA05}.

\begin{figure}[tbh]
\begin{center}
\includegraphics[width=0.5\textwidth]{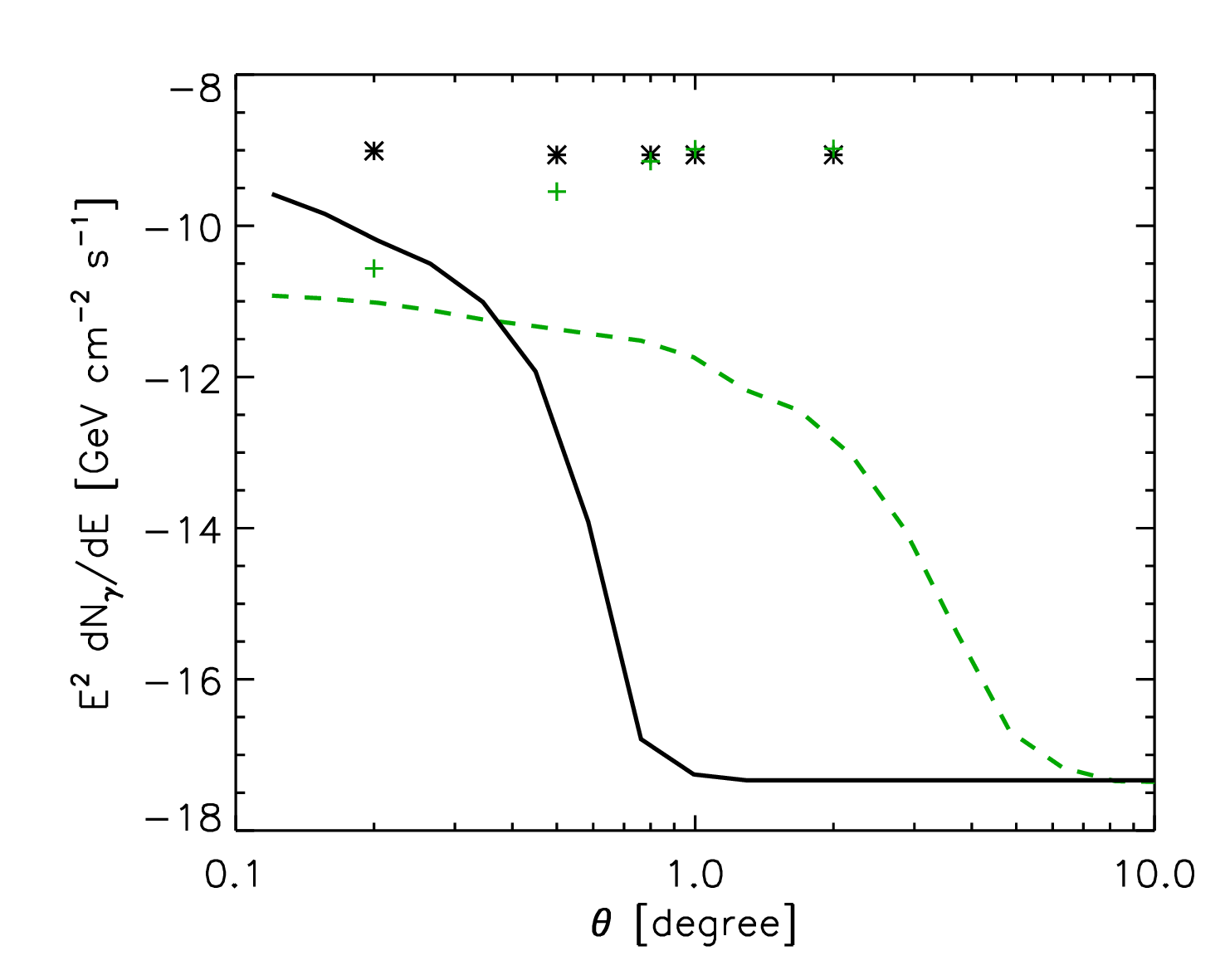} 
\caption{Angular profiles of the images of the sources represented in
  Fig.~\ref{fig:filim}. We represent the gamma ray flux integrated
  over energies $E_\gamma=1-100~$GeV averaged over angular bins, for a
  filament seen along its axis, at 1~Gpc and
  $L_{E,19}=10^{46}$~erg~s$^{-1}$ (black solid line), and at 100~Mpc
  and $L_{E,19}=10^{44}$~erg~s$^{-1}$ (green dashed line). The black
  stars and green crosses present the corresponding integrated flux up
  to a given angular extension in the sky
  $\theta$.} \label{fig:photonprof}
\end{center}
\end{figure}

Figure~\ref{fig:photonprof} demonstrates that the synchrotron gamma
ray emission from both our model sources could be observed by Fermi
LAT and by CTA. Their flux integrated over the angular resolution of
these instruments at $\sim 10~$GeV are indeed above their
corresponding sensitivities. The case presented in dashed green line
is however marginally excluded by the observed spectrum of cosmic rays
and even more so by the search for anisotropy, as discussed previously
(Fig.~\ref{fig:onesource}). It thus appears that the gamma ray flux of
a source of luminosity $L_{E,19}\sim10^{44}$~erg~s$^{-1}$ will be
hardly observable, as placing the source at a further distance to
reconcile its spectrum with the observations will dilute the gamma ray
emission consequently, below the current instrument
sensitivities. More promising are extremely powerful sources with
$L_{E,19}=10^{46}$~erg~s$^{-1}$ located around 1~Gpc.  According to
Fig.~\ref{fig:photonprof}, the emission from such sources would spread
over a fraction of a degree, hence their image could possibly be
resolved by Fermi and certainly by CTA.

As discussed in the introduction, the observation of a multi-GeV
  extended emission as obtained in these images would provide clear
  evidence for acceleration to ultra-high energies, provided the
  possible background associated with a halo seeded by TeV photons can
  be removed. As discussed in ~\cite{ACV94}, \cite{Dai02},
  \cite{dAvezac:2007sg}, \cite{2009PhRvD..80b3010E} and
  \cite{2009PhRvD..80l3012N}, the angular size of the latter signal is
  a rather strong function of the average inter-galactic magnetic
  field. In particular, if the field is assumed homogeneous and of
  strength larger than $10^{-14}\,$G, the image is diluted away to
  large angular scales. Accounting for the inhomogeneity of the
  magnetic field (see Sec.~\ref{sec:compton-cascade} below for a
  detailed discussion), one concludes that the flux associated to this
  image, on a degree scale, is to be reduced by the filling factor of
  voids in which $B\lesssim 10^{-14}\,$G within $\sim 40\,$Mpc of the
  source (this distance corresponding to the energy loss distance of
  the primary $\sim20\,$TeV photons). Furthermore, the angular size of
  the halo seeded by multi-TeV photons depends strongly on photon
  energy, i.e. $\delta\theta\propto E_\gamma^{-2}$ or
  $\delta\theta\propto E_\gamma^{-1.75}$ depending on $\lambda_B$, see
  \cite{2009PhRvD..80l3012N}. In contrast, the synchrotron halo is
  more strongly peaked in energy around $10$\,GeV and in this
  range its angular size is essentially independent of energy, but is
  rather determined by the magnetic field strength and source
  distance. Assuming comparable total power in both signals,
  discrimination should thus be feasible; of course discrimination
  would be all the more easier if the halo signal seeded by multi-TeV
  gamma-rays were diluted to large angular scales.  

The previous discussion thus indicates that detection might be
possible for powerful rare sources at distances in excess of
$100\,$Mpc.  However, one could not then argue, on the basis of the
detection of a gamma ray counterpart that the source of ultrahigh
energy cosmic rays has been identified, as this source of gamma-rays
must remain a rare event, which could not be considered a
representative element of the source population of ultra-high energy
cosmic rays. Overall, one finds that this source must contribute to a
fraction $10\,$\% of the flux at energies well below the GZK cut-off
in order to be detectable in gamma-rays. It is important to note that this 
conclusion does not depend on the degree of beaming of the emission 
of the source, as we discuss in Section~\ref{section:conclusion}.

To conclude this section on a more optimistic note, let us discuss an
effect that could improve the prospects for detection. In the above
discussion and in our simulations, following \cite{GA05}, we have
assumed that all of synchrotron emission takes place in the immediate
environment of the source, as the magnetic field drops to low values
outside the filament, where inverse Compton cascades would therefore
become the dominant source of energy loss of the pairs. However, since
the typical energy loss length of ultrahigh energy cosmic rays exceeds
the typical distance between two filaments, one must in principle
allow for the possibility that the beam of ultrahigh energy protons
seeds secondary pairs in intervening magnetized filaments, which would
then also contribute to the observed synchrotron emission. During the
crossing of a filament, ultrahigh energy protons are deflected by an
angle $\delta\theta_i\,\ll\,1$ (unless the primary cosmic ray carries
a large charge or the filament is of unusual magnetisation, see the
discussion in \citealp{KL08b}). Then, as long as the small deflection
limit applies, the problem is essentially one-dimensional and each
filament contributes to the gamma ray flux a fraction comparable to
that of the source environment. The overall synchrotron signal would
then be amplified by a factor $\lambda/l_{\rm f}$, where $\lambda$
denotes the relevant energy loss length, $\lambda\sim100\,$Mpc for
pion production of $10^{20}\,$eV protons, $\lambda\sim1\,$Gpc for pair
production of $10^{19}\,$eV protons and $l_{\rm f}\sim 40\,$Mpc is the
typical separation between two filaments.  Overall, one might expect
an amplification as large as an order of magnitude for the
contribution to the flux due to pair production, for sources located
at several hundreds of Mpc, and a factor of a few for sources located
at $100\,$Mpc; and, correspondingly, an amplification factor $\sim
2-3$ for the contribution to the flux due to pion production. Since
the pair production channel contributes $\sim10$\% of the total flux
at $E_\gamma\sim 10-100~$GeV for a pure proton composition, the
amplification factor should remain limited to a factor of a few,
unless the interaction length with magnetized regions is substantially
smaller than the above (expected) $l_{\rm f}=40\,$Mpc.

\subsection{Inverse Compton cascades}\label{sec:compton-cascade}

Let us briefly discuss the gamma ray signal expected from Compton
cascades of ultra-high energy photons and pairs injected in the
intergalactic medium. The physics of these cascades has been discussed
in detail in \cite{WTW72}, \cite{Protheroe86}, \cite{PS93},
\cite{ACV94}, \cite{Ferrigno04} and \cite{Gabici07}. The angular
  extent and time distribution of GeV-TeV gamma-rays resulting from
  inverse Compton cascades seeded by ultra-high energy cosmic rays
  produced by gamma-ray bursts have also been discussed by
  \cite{1996ApJ...464L..75W}. Inverse Compton cascades in the steady
state regime have been considered in the study of \cite{ASM06} (for a
source located in a cluster of galaxies) but dismissed in the study of
\cite{GA05} because of the dilution of the emitted flux through the
large deflection of the pairs in the low energy range of the
cascade. Indeed, the effective inverse Compton cooling length of
electrons of energy $E_e\lesssim\,100\,$TeV can be written as
$x_{e\gamma}\,\simeq\, 3.5\,{\rm kpc}\,(E_e/100\,{\rm TeV})^{-1}$ and
on this distance scale, the deflection imparted by a magnetic field of
coherence length $\lambda_B\,\gg\,x_{e\gamma}$ reads
$\theta_e\,\sim\,x_{e\gamma}/r_{\rm L,e}\,\sim \, 3\times 10^{-2}
(E_e/100\,{\rm TeV})^{-2}(B/10^{-12}\,{\rm G})$. Then, assuming that
the last pair of the cascade carries an energy $E_{\rm
  fin}\,\sim\,20\,$TeV (so that the photon produced through the
interaction with the CMB carries a typical energy $\lesssim 1\,$TeV),
one finds that a magnetic field larger than $\sim 10^{-12}\,$G
isotropizes the low energy cascade, in agreement with the estimates of
\cite{GA05}.

  This situation is modified when one takes into account the
  inhomogeneous distribution of extra-galactic magnetic fields, as we
  now discuss. Primary cosmic rays, upon traveling through the voids
  of large scale structure may inject secondary pairs which undergo
  inverse Compton cascades in these unmagnetized regions. If the field
  in such regions is smaller than the above $10^{-12}\,$G, then the
  cascade will transmit its energy in forward $\lesssim $TeV
  photons. Of course, depending on the exact value of $B$ where the
  cascade ends, the resulting image will be spread by some finite
  angle. Since we are interested in sharply peaked images, let us
  consider a typical angular size $\theta$ and ignore those regions in
  which the magnetic field is large enough to give a contribution to
  the image on a size larger than $\theta$. For $\theta\ll1$, the
  problem remains one-dimensional as before, and one can compute the
  total energy injected in inverse Compton cascades within $\theta$,
  as follows.

  The luminosity injected in secondary pairs and photons up to
  distance $d$ is written $\chi_e L_{\rm cr}(>E)$. Since we are interested
  in the signatures of ultrahigh cosmic ray sources, we require that
  $E\geq10^{19}\,$eV; for protons, the energy loss length due pair
  production moreover increases dramatically as $E$ becomes smaller
  than $10^{19}\,$eV, so that the contribution of lower energy
  particles can be neglected in a first approximation. For photo-pair
  production, the fraction transfered is $\chi_{e,ee}\simeq d/1\,{\rm
    Gpc}$ of $L_{E,19}=L_{\rm cr}(>10^{19}\,{\rm eV})$ up to $d\sim1\,$Gpc. For
  pion production, the fraction of energy transfered is roughly
  $\chi_{e,\pi}\simeq d/100\,{\rm Mpc}$ of $L_{\rm
    cr}(>6\,10^{19}\,{\rm eV})$ in the continuous energy loss
  approximation. At distances $100\,{\rm Mpc}\,\leq\,d\,\leq\,1\,{\rm
    Gpc}$, the fraction $\chi_e$ of $L_{E,19}$
  injected into secondary pairs and photons thus ranges from $\sim
  0.5$ for $d=100\,$Mpc to $\sim1$ at $d=1\,$Gpc; in short, it is
  expected to be of order unity or slightly less.  All the energy
  injected in this way in sufficiently unmagnetized regions (see
  below) will be deposited through the inverse Compton cascade in the
  sub-TeV range, with a typical energy flux dependence $\propto
  E_\gamma^{1/2}$ up to some maximal energy $E_{\gamma,\rm max}\sim
  1-10\,$TeV beyond which the Universe is opaque to gamma rays on the
  distance scale $d$ \citep{Ferrigno04}. Neglecting any redshift
  dependence for simplicity, the gamma-ray energy flux per unit energy
  interval may then be approximated as:
\begin{eqnarray}
E_\gamma^2 \frac{{\rm d}N_\gamma}{{\rm d}E_\gamma}&\,\approx\,&
 f_{\rm 1d}(<B_\theta)\,\chi_e\frac{L_{\rm cr}}{8\pi
    d^2}\left(\frac{E_\gamma}{E_{\gamma,\rm
        max}}\right)^{1/2}\nonumber\\
& \,\simeq\,& 2.5\times10^{-10}\,{\rm GeV}\,{\rm cm}^{-2}\,{\rm
  s}^{-1}\,f_{\rm 1d}(<B_\theta)\chi_e\nonumber\\
& & \quad\quad\times\frac{L_{E,19}}{10^{42}\,{\rm erg/s}} 
\left(\frac{d}{100\,{\rm Mpc}}\right)^{-2}\left(\frac{E_\gamma}{E_{\gamma,\rm
        max}}\right)^{1/2}\ .
\end{eqnarray}
where $f_{\rm 1d}(<B_\theta)$ denotes the one-dimensional filling
factor, i.e. the fraction of the line of sight in which the magnetic
field is smaller than the value $B_\theta$ such that the deflection of
the low energy cascade is $\theta$. For reference, $B_\theta
\,\sim\,2\times 10^{-14}\,$G for $\theta\sim1^\circ$. In general, one
finds in the literature the three-dimensional filling factor $f_{\rm
  3d}$, but $f_{\rm 1d}(<B_\theta) \sim f_{\rm 3d}(<B_\theta)$ up to a
numerical prefactor of order unity that depends on the geometry of the
structures. Interestingly enough, the amount of magnetization of the
voids of large scale structure is directly related to the origin of
large scale magnetic fields. Obviously, if galactic and cluster
magnetic fields originate from a seed field produced in a homogeneous
way with a present day strength $B\gg 10^{-14}\,$G, then the above
gamma ray flux will be diluted to large angular scales, hence below
detection threshold. However, if the seed field, extrapolated to
present day values is much lower than this value, or if most of the
magnetic enrichment of the intergalactic medium results from the
pollution by star forming galaxies and radio-galaxies, then one should
expect $f_{\rm 1d}(<B_\theta)$ to be non negligible. For instance,
\cite{2009MNRAS.392.1008D} obtain $f_{\rm 3d}(<10^{-14}\,{\rm G})\sim
{\cal O}(1)$ in such models. Given the sensitivity of current gamma ray
experiments such as H.E.S.S. in the TeV energy range, the inverse
Compton cascades might then produce degree-size detectable halos for
source luminosities $\gtrsim 10^{43}[f_{\rm 3d}(<10^{-14}\,{\rm G})]^{-1}(d/100\,{\rm
  Mpc})^{-2}\,$erg/s. For future instruments such as CTA, the
detectability condition will be of order $L_{\rm
  s}\gtrsim10^{41}[f_{\rm 3d}(<10^{-14}\,{\rm G})]^{-1}(d/100\,{\rm Mpc})^{-2}\,$erg/s. 
  The expected flux level remains quite uncertain as it depends mostly 
  on the configuration of the extragalactic magnetic field in the voids, 
  contrarily to the synchrotron signal which is mainly controlled by the source luminosity. 
  
  We note that
intergalactic magnetic fields of strength $B\lesssim10^{-15}\,$G might
be probed through the delay time of the high energy afterglow of
gamma-ray bursts \citep{1995Natur.374..430P,2008ApJ...682..127I} or
the GeV emission around blazars
\citep{ACV94,Dai02,dAvezac:2007sg,2009PhRvD..80b3010E,2009PhRvD..80l3012N},
as discussed earlier. As the present paper was being refereed, some
 first estimates on lower bounds for the average magnetic field appeared, 
 giving $B\gtrsim 10^{-16}-10^{-15}$~G, see \cite{Neronov10}, \cite{Ando10} and \cite{Dolag10}.

\subsection{Cen A}

In view of the results of section~\ref{subsection:average}, one may
want to consider also the case of mildly powerful but nearby
sources. One such potential source is Cen~A, which has attracted a
considerable amount of attention, as it is the nearest radio-galaxy
(3.8~Mpc) and more recently, because a fraction of the Pierre Auger
events above $60\,$EeV have clustered within $10-20^\circ$ of this
source. As discussed in detail in \cite{LW09}, this source is likely
too weak to accelerate protons to $\gtrsim 10^{19}\,$eV in steady
state, but heavy nuclei might possibly be accelerated up to GZK
energies. The acceleration of protons to ultrahigh energies in flaring
episodes of high luminosity has been proposed in \cite{DA09}. It is
also important to recall that the apparent clustering in this
direction can be naturally explained thanks to the large concentration
of matter in the local Universe, in the direction of Cen~A but located
further away. As discussed in \cite{LW09}, some of the events detected
toward Cen~A could also be attributed to rare and powerful bursting
sources located in the Cen~A host galaxy, such as gamma-ray
bursts. Due to the scattering of the particles on the magnetized
lobes, one would not be able to distinguish such bursting sources from
a source located in the core of Cen~A.

We have argued throughout this paper that a clear signature of
ultrahigh energy cosmic ray propagation could be produced in presence
of an extended and strong magnetized region around the source. The
magnetized lobes of Cen A provide such a site, with an extension of
$R_{\rm lobe}\sim 100~$kpc, and a magnetic field of average intensity
$B_{\rm lobe}\sim 1~\mu$G and coherence length $\lambda_{\rm lobe}\sim
20~$kpc \citep{Feain09}.  In terms of gamma ray fluxes, one can calculate however that
the only advantage Cen A presents compared to average types of sources
described in section~\ref{subsection:average} lies in its
proximity. As noted before, the bulk of gamma ray emission will be due
to electrons and positrons produced by pion production with $E_e >
10^{19}$~eV. The flux of these secondary electron/positron emission
scales as the distance $r$ the primary protons travels through, as
long as $r\ll \lambda$, with $\lambda$ the energy loss length by pion
production.  In our situation, in order to calculate the synchrotron
gamma ray flux, this distance corresponds to the size of the
magnetized structure with $B\gtrsim 0.5\,\mbox{nG} \,
(E_e/10^{19}\,{\rm eV})^{-3/4}$, namely $r\sim R_{\rm lobe}$ in our
toy-model. One may also note that considering the magnetic field
strength in the lobes, the gamma ray flux will peak around $E_{\gamma
  \rm syn} \sim 6.8\,$TeV$\,B_{\rm lobe}E_{e,19}^2$, according to
Eq.~\ref{eq:Esyn}. Scaling the expected flux from the lobes around
$10~$TeV $F_{\rm lobe,10\,TeV}$ to the flux that we calculated in
section~\ref{subsection:average} for standard sources embedded in
filaments around $10~$GeV $F_{\rm fil,10\,GeV}$, one obtains:
\begin{eqnarray}
F_{\rm lobe,10\,TeV} &\sim& \left(\frac{d_{\rm Cen\,A}}{d_{\rm fil}}\right)^{-2}\,
\frac{L_{\rm CenA}}{10^{42}\,{\rm erg/s}}\,\frac{R_{\rm lobe}}{5\,{\rm
  Mpc}}\, F_{\rm fil,10\,GeV}\\
&\sim& 2\times 10^{-14} ~\mbox{GeV}~\mbox{s}^{-1} \mbox{cm}^{-2}\, ,
\end{eqnarray}
which is far below current instrument sensitivities.  
Here, $d_{\rm  CenA}=3.8~$Mpc 
and $d_{\rm fil}=100$~Mpc are the distance of Cen~A and of our model filament
respectively, 
and $L_{\rm CenA}$ is the cosmic ray luminosity
of Cen~A which we set to $10^{39}$~erg~s$^{-1}$ according to
constraints from ultrahigh energy cosmic ray observations (see for
example \citealp{Hardcastle09}). 
Moreover, Cen A covers
$\theta_{\rm source}\sim10^\circ$ in the sky, inducing a sensitivity
loss of a factor $\theta_{\rm source}/\theta_{\rm PSF}\sim 10$ for the
Fermi telescope ($\theta_{\rm PSF}$ represents the angular resolution
of the instrument). 

All in all, taking into account all these disadvantages, we conclude
that the synchrotron emission produced in the lobes of Cen A would be
three orders of magnitudes weaker than the flux obtained for the
average types of sources described in
section~\ref{subsection:average}.

Our numerical estimate using the method introduced in
section~\ref{subsubsection:method} and modeling the lobes as spherical
uniformly magnetized structures confirms this calculation. We note that the contribution of
the pair production channel is not as suppressed as the other
components. Due to the strength of the magnetic field, the electrons
and positrons radiating in synchrotron around $0.1-1~$GeV are of lower
energy than for the standard filament case ($E_e\sim 10~$MeV) and thus
are more numerous. At these energies, the synchrotron emission in the
filament was also partially suppressed by the shortening of the
inverse Compton loss length as compared to the synchrotron loss
length. Because of the high magnetic field intensity, this effect is
less pronounced in the lobe case.

Recently Fermi LAT discovered a gamma ray emission emanating from the
giant radio lobes of Cen A \citep{Cheung10}. In light of our
discussion, the ultrahigh energy cosmic ray origin of this emission
can be excluded. It could rather be interpreted as an inverse Compton
scattered relic radiation on the CMB and infrared and optical photon
backgrounds, as proposed by \cite{Cheung10}.

Let us note additionally that \cite{KOT09} have provided estimates for
gamma ray fluxes emitted as a result of the production of ultrahigh
energy cosmic rays in Cen~A. These authors conclude that a substantial
signal could be produced, well within the detection capabilities of
Fermi and CTA. However this signal results from interactions of cosmic
rays accelerated inside the source, not outside. As we discussed
before, such a gamma ray signal, i.e. that could not be angularly
resolved from the source, could not be interpreted without
ambiguity. In particular, one would not be able to discriminate it
from a secondary signal produced by multi-TeV leptons or hadrons
accelerated in the same source. Furthermore, the gamma ray signal
discussed in \cite{KOT09} is mainly attributed to cosmic rays at
energies $\lesssim 10^{18}\,$eV, not to ultrahigh cosmic rays above
the ankle, although the generation spectrum in the source is
normalized to the Pierre Auger Observatory events at $\gtrsim 6 \times
10^{19}\,$eV. The conclusions thus depend rather strongly on the shape
of the injection spectrum as well as on the phenomenological modelling
of acceleration, as made clear by Fig.~1 of their study.

Finally one may turn to the highest energies and search for ultrahigh
energy cosmic ray signatures from Cen~A by looking directly at the
ultrahigh energy secondary photons produced during the
propagation. These photons should indeed not experience cascading at
this close distance. \cite{Taylor09} have studied this potential
signature and concluded that Auger should be able to detect
$0.05-0.075$~photon per year from Cen~A, assuming that it is
responsible for 10\% of the $>6\times 10^{19}$~eV cosmic ray flux, and
assuming a 25\% efficiency for photon discrimination.

\section{Discussion and conclusions}\label{section:conclusion}
Let us first recap our main findings. We have discussed the
detectability of gamma ray halos of ultrahigh energy cosmic ray
sources, relaxing most of the assumptions made in previous studies.
We have focused our study on the synchrotron signal emitted by
secondary pairs, which offers a possibility of unambiguous detection.
In this study, we have taken into account the inhomogeneous
distribution of the magnetic field in the source environment and
examined the effects of non pure proton composition on the gamma ray
signal.  We find that the gamma ray flux predictions are rather robust
with respect to the uncertainties attached to these physical
parameters. The normalization and hence the detectability of this flux
ultimately depends on the energy injected in the primary cosmic rays,
for realistic values of the magnetic field.  We have further
demonstrated that the average type of sources contributing to the
ultrahigh energy cosmic ray spectrum produces a gamma ray flux more
than two orders of magnitudes lower than the sensitivity of the
current and upcoming instruments. The case of rare powerful sources
contributing to more than 10\% of the cosmic ray flux is far more promising in terms of
detectability. We have found that gamma ray signatures of those
sources could be detectable provided that they are located far enough
not to overshoot the observed cosmic ray spectrum. If the extended
emission of such signatures were resolved (which should be the case
with Fermi and CTA), such a detection would provide a distinctive
proof of acceleration of cosmic rays to energies $\gtrsim
10^{19}\,$eV. However, as we have discussed, such gamma ray sources
should remain exceptional and could not be considered as
representatives of the population of ultrahigh energy cosmic ray
sources. We have also discussed the deflection and the dilution of the
Compton cascading gamma ray signal at sub-TeV energies. In this case,
the flux prediction depends sensitively on the filling factor of the
magnetic field with values smaller smaller than $\sim 10^{-14}\,$G
in the voids of large scale structure. If this latter is larger than a
few percents, as suggested by some scenarios of the origin of cosmic
magnetic fields, then the inverse Compton cascade could provide a
degree size image of the ultra-high cosmic ray source, with a flux
slightly larger than that associated to the synchrotron component.

Finally, we have also discussed the detection of nearby sources,
considering the radiogalaxy Centaurus~A as a protoypical example.  We
have demonstrated that, {\em if} ultrahigh energy cosmic rays are
produced in Cen~A, their secondary gamma ray halo signature should not
be detectable by current experiments. Our conclusion indicates in
particular that the gamma ray emission detected by Fermi LAT from the
giant radio lobes of Cen A \citep{Cheung10} cannot result from the
synchrotron radiation of ultrahigh energy secondaries in that source.

We need to stress that, in our discussion, we have systematically
assumed that the source was emitting cosmic rays at a steady rate with
isotropic output. Let us discuss these assumptions. If the source is
rather of the flaring type, then, at a same cosmic ray source
luminosity, the flux predictions will be reduced by the ratio of the
source flare duration $\Delta t$ to the dispersion in arrival times
$\sigma_t$ that results from propagation in intergalactic magnetic
fields. In theoretical models, $\Delta t$ can take values ranging from
a few seconds for gamma-ray
bursts~\citep{1996APh.....4..365M,1995PhRvL..75..386W,1995ApJ...453..883V}
to days in the case of blazar flares, while $\sigma_t$ is typically
expected to be of the order of $10^3-10^5\,$yrs for protons of energy
$\gtrsim 70\,$EeV~(see \citealp{KL08b} for a detailed discussion). In
both cases, the large luminosity during the flaring episodes is not
sufficient to compensate for the dilution in arrival times. Of course,
if the source is a repeater, with a timescale between two flaring
episodes shorter than the intergalactic dispersion timescale, then one
recovers a steady emitting situation, with an effective luminosity
reduced by the fraction of time during which the source is
active. 

This latter situation may well apply to powerful radio-galaxies
  that produce ultra-high energy cosmic rays with flaring activity,
  since the size and magnetization of the jets and lobes are
  sufficient to deflect ultra-high energy cosmic rays by an angle of
  order unity, theredy dispersing them in time over $\sim R_{\rm
    lobe}/c\sim 3\times 10^5\,$yr, with $R_{\rm lobe}\sim 100\,$kpc
  the typical transverse scale of the lobes. Even if ultra-high energy
  cosmic rays are ejected as neutrons, as discussed for instance by
  \cite{DA09}, a substantial fraction of these cosmic rays will suffer
  dispersion as the decay length of neutrons of energy $E_n$ is
  $0.9(E_n/10^{20}\,{\rm eV})\,$Mpc. In such a configuration, the
  source would effectively emit its cosmic rays isotropically, as we
  have assumed.

  Let us nonetheless discuss the case of beamed sources, for the sake
  of completeness. If one considers a single source, emitting cosmic
  rays in a solid angle $\Delta \Omega_{\rm s}< 4\pi$, then at a same
  total luminosity as an isotropic source, the luminosity emitted per
  steradian is larger by $4\pi/\Delta\Omega_{\rm s}$ and so is the
  expected gamma-ray flux (in the limit of small deflection around the
  source, which is necessary for detection). In other words, the
  required total cosmic-ray luminosity per source for detection of
  gamma rays is lowered by a factor $\Delta\Omega_{\rm
    s}/(4\pi)$. However, at a fixed contribution of the source to the cosmic ray flux, 
    the expected gamma-ray flux does not depend on the beaming angle. 
    This is because both the cosmic ray flux and the gamma-ray flux are 
    determined by the number of visible sources times the luminosity
  per steradian\footnote{The amount of magnetic deflection suffered
    during propagation in extra-galactic magnetic fields does not
    affect this result, as the larger number of visible sources for a
    larger amount of deflection is compensated by the dilution of the
    flux of each source.}. In particular, our conclusion that a source must 
    contribute to more than 10\% of the ultra-high energy cosmic ray flux 
    in order for the synchrotron signal to be detectable remains unchanged
     if $\Delta\Omega_{\rm s}<4\pi$.

\section{Acknowledgement}
We thank Eric Armengaud, Stefano Gabici, Susumu Inoue, Simon Prunet
and Michael Punch for helpful discussions. We thank Christophe Pichon
for helping us identify adequate filaments in our cosmological
simulation cube using his skeleton program, and St\'ephane Colombi for providing us with the Dark Matter cosmological simulation cube. KK is supported by the NSF
grant PHY-0758017 and by Kavli Institute for Cosmological Physics at
the University of Chicago through grant NSF PHY-0551142 and an
endowment from the Kavli Foundation.

\bibliographystyle{aa} 
\bibliography{KAL_v2} 

\end{document}